\documentclass{aastex631}

\usepackage{longtable}
\usepackage{rotating}
\usepackage{epsfig}
\usepackage{amsmath}
\usepackage{threeparttable}

\begin{document}
\title{Evidence for the transition from thermal to non-thermal emission in the prompt emission of GRB 161117A}

\author[0000-0001-8217-8889]{Xue-Zhao Chang}
\affiliation{Institute of Astrophysics, Chuxiong Normal University, Chuxiong 675000, China}
\affiliation{Guangxi Key Laboratory for
Relativistic Astrophysics, School of Physical Science and Technology, Guangxi University,
Nanning 530004, China; lhj@gxu.edu.cn}
\author[0000-0001-6396-9386]{HouJun L\"{u}} 
\altaffiliation{Corresponding author (LHJ) email: lhj@gxu.edu.cn}
\affiliation{Guangxi Key Laboratory for
Relativistic Astrophysics, School of Physical Science and Technology, Guangxi University,
Nanning 530004, China; lhj@gxu.edu.cn}
\author[0000-0001-5681-6939]{Jia-Ming Chen}
\affiliation{Department of Astronomy, School of Physics and Astronomy, Yunnan University, Kunming 650091, China}
\author[0000-0002-7044-733X]{En-Wei Liang}
\affiliation{Guangxi Key Laboratory for
Relativistic Astrophysics, School of Physical Science and Technology, Guangxi University,
Nanning 530004, China; lhj@gxu.edu.cn}

\begin{abstract}
GRB 161117A is a long-duration GRB with three main overlapping peaks. By analyzing the time-resolved spectra of its data observed with the Gamma-Ray Burst Monitor (GBM) on board the Fermi mission, we find that the spectral evolution shows a transition from thermal (single BB) to hybrid (PL$+$BB), and finally to non-thermal (Band and CPL) emissions. Such a transition suggests that the jet composition of GRB 161117A should be changed from a fireball to a Poynting-flux-dominated jet. The bulk Lorentz factor ($\Gamma_{\rm ph}$), radii ($R_{\rm ph}$ and $R_{0}$), magnetization factor at the central engine ($\sigma_0$), and dimensionless entropy ($\eta$) of the outflow can be inferred by invoking the observed quasi-thermal component within two models (e.g., pure fireball and hybrid). It is found that $\Gamma_{\rm ph}$ seems to be tracking with the light curve, and $R_{0}$ remains a constant at $\sim$ $10^{8}$ cm. The low magnetization ($1+\sigma_0 \sim$ 1) and high dimensionless entropy ($\eta \gg$ 1) during the first seven time-intervals suggest to be a pure fireball outflow. Moreover, we also estimate the lower limit of magnetization parameter at the photosphere radius ($\sigma_{\rm ph}\sim 1.4$ and 0.75) for late phase via the non-thermal spectra, and it indicates that the particle acceleration mechanism is dominated by internal shocks rather than magnetic dissipation processes. Finally, the $\nu \bar{\nu}$ annihilation mechanism of NDAF model to explain the thermal emission of GRB 161117A is also discussed.

\end{abstract}

\section{Introduction} \label{sec:intro}
Gamma-ray bursts (GRBs) are among the most intense and energetic phenomena observed in the universe. Despite significant advancements in both observation and theory over the past 50 years \citep{2015PhR...561....1K}, several key questions remain unresolved, such as the nature of the prompt emission mechanism and the composition of jets \citep{2011CRPhy..12..206Z,2018pgrb.book.....Z,2019MmSAI..90...57M}.

The fireball model was widely adopted to interpret the observations in GRB studies, and the GRB jet initially forms as a hot fireball which is composed of photons, electron-positron pairs, and a small fraction of baryons \citep{1986ApJ...308L..47G,1986ApJ...308L..43P,1990ApJ...365L..55S,1998ApJ...494L..45P}. The optical depth at the initial radius of the fireball $R_{0}$ is much larger than unity, and the thermal photons escape around the fireball photosphere $R_{ph}$ where the optical depth $\tau \sim 1$. Then, it can produce the observed thermal emission (photosphere model; \citealt{1986ApJ...308L..47G,1986ApJ...308L..43P,1990ApJ...365L..55S}). After a rapid acceleration phase driven by thermal pressure, a fraction of the thermal energy is converted into the kinetic energy of the outflow \citep{1993ApJ...415..181M}. Meanwhile, the expanding baryonic matter dissipates the kinetic energy through internal collisions to produce energetic particles. These particles then are emitted via synchrotron (or synchrotron self-Compton) radiation to generate the observed non-thermal $\gamma$-ray emission (internal shock model; \citealt{1994ApJ...430L..93R,1997ApJ...490...92K,1998MNRAS.296..275D,2014AcASn..55..354Z}). Another scenario invokes a non-thermal component from the Poynting-flux-dominated outflow where most of the energy is stored in the magnetic field \citep{2011ApJ...726...90Z}. The magnetic energy can be dissipated through magnetic reconnection or current instability to power the observed prompt emission of GRB \citep{Thompson1994,2003astro.ph.12347L,2009ApJ...700L..65Z}, or the internal-collision-induced magnetic reconnection and turbulence (ICMART) model \citep{2011ApJ...726...90Z}. 

Observationally, the spectra of a larger fraction of GRB prompt emission is dominated by a purely non-thermal component, such as GRB 080916C \citep{2009Sci...323.1688A,2009ApJ...700L..65Z}, GRB 130606B \citep{2016ApJ...816...72Z}, and GRB 230307A \citep{2024MNRAS.529L..67D,2023arXiv231007205Y}. On the contrary, the spectrum of GRB 090902B is dominated by a purely quasi-thermal component \citep{2009ApJ...706L.138A,2010ApJ...709L.172R}. By comparing with the non-thermal spectrum, the thermal spectrum is much narrower, and is much harder at the low-energy band. In general, the dissipated radius of the photospheric and internal shock cannot be distinguished very well, and the observed spectrum of GRB prompt emission should be the superposition of thermal and non-thermal components \citep{2005ApJ...625L..95R,2015ApJ...801...2}. This thermal component remains weak if the initial magnetization is large enough \citep{2002MNRAS.336.1271D, 2013A&A...551A.124H}. A subdominant blackbody (BB) component that is superposition on the non-thermal component has been observed in a number of GRBs, e.g., GRB 081221 \citep{2018ApJ...866...13H}; GRB 100724B \citep{2011ApJ...727L..33G}; GRB 110721A \citep{2012ApJ...757L..31A}; GBR 160625B \citep{2017ApJ...849...71L,2018NatAs...2..258Z}; 210610B \citep{2022ApJ...932...25C}; and GRB 211211A \citep{2023ApJ...943..146C}.

In terms of theoretical models, one of the competitive models of the GRB central engine is a rotating black hole (BH) surrounded by a neutrino-dominated accretion flow (NDAF). Two mechanisms are considered to power the relativistic jet in a GRB for a BH central engine: one is the neutrino-antineutrino ($\nu \bar{\nu}$) annihilation mechanism which releases energy from the accretion disk \citep{1999ApJ...518..356P,2002ApJ...579..706D,2006ApJ...643L..87G,2007ApJ...657..383C,2007ApJ...664.1011J,2009ApJ...700.1970L,2015ApJS..218...12L}. and the other one is the Blandford-Znajek (BZ; \citealt{1977MNRAS.179..433B}) mechanism which extracts the spin energy from the Kerr BH \citep{2000PhR...325...83L,2000PhRvD..61h4016L,2013ApJ...765..125L}. These two processes may operate simultaneously in a hyper-accretion system. Recent studies have suggested that the thermal component in GRB spectra may originate from $\nu \bar{\nu}$ annihilation \citep{2024ApJ...975..225L}, while other work proposed that the central engine may first emit a hot fireball via $\nu \bar{\nu}$ annihilation, followed by a Poynting-flux-dominated jet via the BZ mechanism \citep{2017ApJ...849...47L}.

An interesting question is whether the jet composition can be changed between pulses of prompt emission in a single GRB. For example, GRB 160625B shows a transition from a fireball to a Poynting-flux-dominated outflow from early precursor emission to late main bursts \citep{2017ApJ...849...71L,2018NatAs...2..258Z,2019ApJS..242...16L}.

Recently, by analysing the spectra of GRBs observed by Fermi/Gamma-ray Burst Monitor (GBM), it is found that GRB 161117A exhibited three distinct pulses, characterized by a low-energy spectral index ($\alpha$) that exceeded the ``death line'' of synchrotron emission\footnote{In the slow-cooling synchrotron emission scenario, the maximum value of $\alpha$ cannot exceed -2/3, which is referred to as the ``death line of synchrotron emission'' \citep{1998ApJ...506L..23P}.}. Additionally, the spectra displayed an evolution with a hard-to-soft trend. In this paper, we employ a Bayesian method to analyze the time-resolved spectra \footnote{Due to the potential evolution of jet components, this paper focuses solely on the analysis of time-resolved spectra.} of GRB 161117A in order to investigate its jet components, and find that the spectral evolution presents a transition from thermal to non-thermal emission. The structure of this paper is organized as follows. The data analysis and spectral fitting are presented in Section 2. In Section 3, we derive the fireball parameters for the early thermal phase, and check the Poynting-flux-dominated outflow for the later non-thermal phase in Section 4. In Section 5, we adopt the GRB quasi-thermal emission to test the NDAF model under the fireball assumption. The conclusions are drawn in Section 6 with some additional discussion. Throughout the paper, a concordant cosmology with parameters $H_{0}=70 \mathrm{~km} \mathrm{~s}^{-1} \mathrm{Mpc}^{-1}$, $\Omega_{\mathrm{M}}=0.30$ and $\Omega_{\Lambda}=0.70$ is adopted and we use the notation $Q=10^{n} Q_{n}$ in CGS units.

\section {Data Analysis}
GRB 161117A (trigger 501039335/161117066) triggered was Fermi/GBM at 01:35:31.36 UT on November 17, 2016 \citep{2016GCN.20192....1M}, and also was observed by several other instruments, including Swift/BAT \citep{2016GCN.20188....1S}, Swift/XRT \citep{2016GCN.20185....1P}, Swift/UVOT \citep{2016GCN.20190....1M}, and the TAROT La Silla observatory \citep{2016GCN.20194....1K}. The redshift of the event was determined to be approximately $z=1.549$ via VLT/X-shooter \citep{2016GCN.20180....1M}. 

\subsection{Light Curve Extraction}
\begin{figure}
\centering
 \includegraphics [angle=0,scale=0.5] {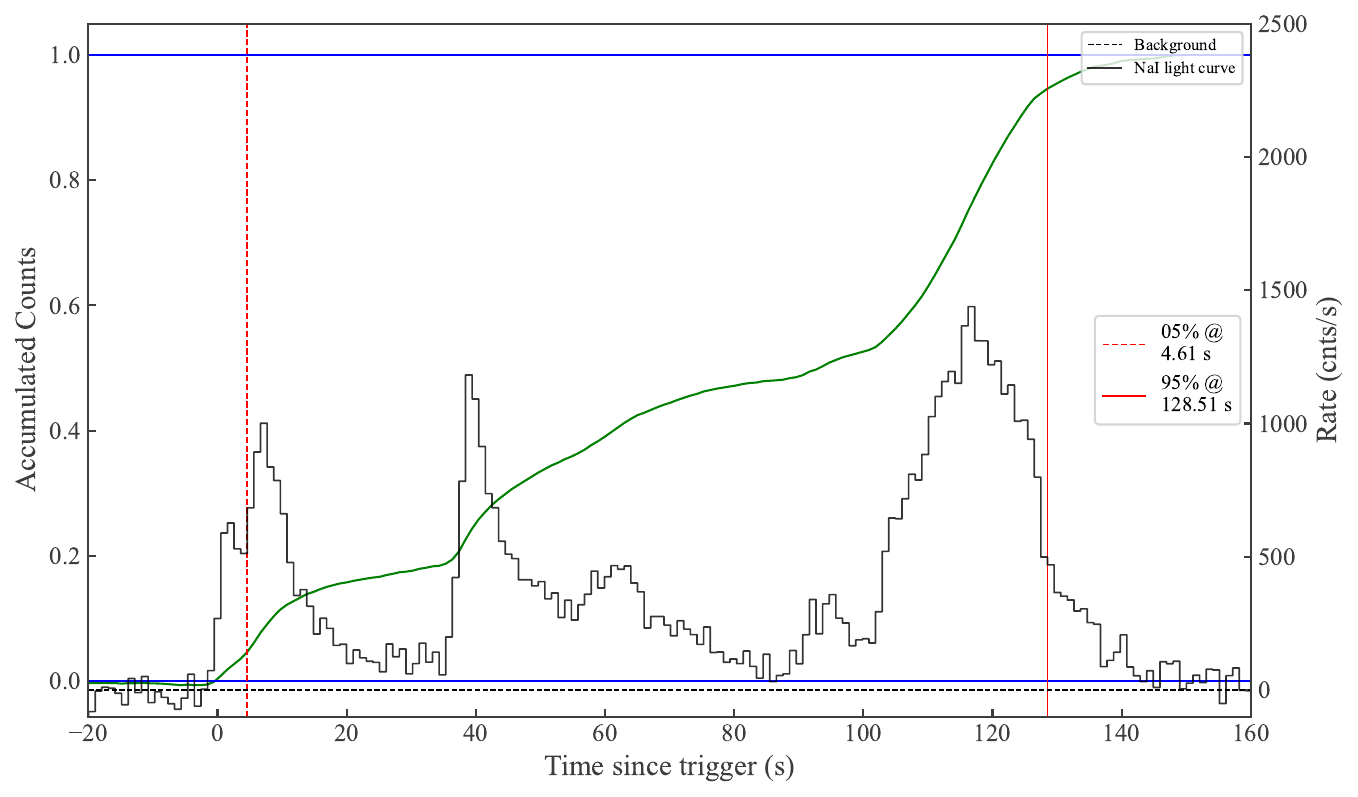}
 \caption{Fermi/GBM light curve of GRB 161117A and the determination of its $T_{90}$. The left y-axis is the photon cumulative counts of the light curve, while the right y-axis is the count rate of the light curve. The solid black line corresponds to the light curve of GRB 161117A, and the dashed black horizontal line is the background. The red dashed and solid vertical lines are the lower and upper limits of $T_{90}$, respectively. The solid blue horizontal lines mark the 0 and 1 photon cumulative counts.}
 \label{fig:1}
\end{figure}

The Fermi/GBM comprises 12 sodium iodide (NaI) detectors and 2 bismuth germanate (BGO) detectors \citep{2009ApJ...702..791M}. The NaI detectors span the energy range from 8 keV to 1 MeV, while the BGO detectors cover energies from 200 keV to 40 MeV. We download the corresponding time-tagged-event (TTE) data for GRB 161117A from the public science support center on the official Fermi website \footnote{https://heasarc.gsfc.nasa.gov/FTP/fermi/data/gbm/daily/}. For the analysis, we select the two brightest NaI detectors and one BGO detector (n1, n2, and b0). The light curves are extracted using a time bin of 1.024 s (see Figure \ref{fig:1}) by running the spectrum source package $Fermi ScienceTools$. Figure \ref{fig:1} shows the extracted light curve of GRB 161117A, which exhibits three main overlapping peaks with increasing amplitude, superimposed with numerous rapid variations. The duration of GRB 161117A, called $T_{90}$, is calculated for NaI detectors within the energy range of 8 - 900 keV as $T_{90}\sim 124$ s (see Figure \ref{fig:1}).

\subsection{Spectral Fits and Results}
We extract the raw photon count spectrum from the TTE data for 128 energy channels. The background for each detector is modeled using polynomial fitting with polynomials ranging from first to fourth order fitted to intervals before and after the burst, and the background model is subtracted from the observed data. The selected energy range approximately spans from 8 keV to 900 keV and 250 keV to 40 MeV for the NaI and BGO detectors, respectively. The energy range from 30 keV to 40 keV is excluded in our analysis due to the K-edge at 33.17 keV caused by the instrument itself. The standard response file (RSP) provided by the GBM team is performed to do the spectral fitting process. Maximum-likelihood-based statistics, called PGstat, are used with a Poisson likelihood and Gaussian likelihood for observation and background, respectively \citep{1979ApJ...228..939C}.

For the subsequent time-resolved spectral analysis of GRB 161117A, we choose the brightest detector (e.g., n2) among NaI and BGO to do the analysis, that is because the brightest detector has a minimum angle between the incident photon and the normal direction of the detector. In order to analyze the spectral evolution, the signal-to-noise ratio (S/N) for each time-bin should be high enough \citep{2018ApJS..236...17V}. So, each broad pulse of the prompt emission is divided into three episodes: pre-peak, around the peak, and post-peak. Each time-bin is checked to ensure a S/N$\geq$ 15 with sufficient photons for reliable spectral fitting. The time-resolved spectral analysis of GRB 161117A is conducted over the interval from $T_0$ to $T_0+$138 s, and separate to nine time slices (named as from 1st to 9th): [0, 4.6] s, [4.6, 8.6] s, [8.6, 30.0] s, [30.0, 38.0] s, [38.0, 42.0] s, [42.0, 85.0] s, [85.0, 115.0] s, [115.0, 125.0] s, and [125.0, 138.0] s, as shown in Table \ref{table1}.

In our analysis, several spectral models can be selected to do the both time-integrated and time-resolved spectral fitting of the burst, such as (1) Band function (Band), (2) cutoff power-law (CPL), (3) blackbody (BB), (4) power-law (PL) models, as well as combinations of any two models. Those models are adopted to do the fits of spectra in GRB studies \citep{2011ApJ...730..141Z,2015ApJ...814...10G}. The details of above models are shown as follows:

(1). The Band function \citep{1993ApJ...413..281B} model can be written as,
 \begin{eqnarray}
N_{\textrm{Band}}(E)=A\left\{\begin{array}{clcc}
(\frac{E}{100~\mathrm{keV}})^{\alpha }\mathrm{exp}\left[-\frac{(\alpha+2)E}{E_{p}} \right ], E \leq (\alpha-\beta) E_{c}, \\
(\frac{E}{100~\mathrm{keV}})^{\beta }\mathrm{exp}(\beta -\alpha )(\frac{(\alpha-\beta)E_{c}}{100~\mathrm{keV}})^{\alpha-\beta }, E > (\alpha-\beta) E_{c}
\end{array}\right.
\end{eqnarray}
where $A$ is the normalization of the spectrum, $\alpha$ and $\beta$ are the low and high-energy photon spectral indices, respectively. $E_{\rm p}=(2+\alpha)E_{\rm c}$ is the peak energy and $E_{\rm c}$ is the cut-off energy.

(2). The CPL model is expressed as,
\begin{eqnarray}
N_{\rm CPL}(E) = A\cdot (\frac{E}{100~\mathrm{keV}})^{\alpha} \mathrm{exp}(-\frac{E}{E_{c}}).
\end{eqnarray}

(3). A Blackbody (BB) emission of quasi-thermal component \citep{,2010ApJ...709L.172R} can be expressed as Planck function,
\begin{equation}
N_{\textrm{BB}}(E)=A\cdot \frac{E^{2}}{\exp [E / k T]-1}
\end{equation}
where $k$ and $T$ are Boltzmann constant and temperature, respectively.

(4). A power-law (PL) component is written as
\begin{equation}
N_{\textrm{PL}}(E)=A\cdot (\frac{E}{100~\mathrm{keV}}) ^{\gamma}
\end{equation}
where $\gamma$ is the photon index of spectrum. We perform spectral fitting by adopting the Multi-Mission Maximum Likelihood (3ML) package \citep{2015arXiv150708343V}, which employs the Markov Chain Monte Carlo (MCMC) method for time-resolved spectral fitting. More details of spectral fitting can refer to our previous paper of \citealt{2024ApJS..275....9C}.


In order to select the most optimal model from the set of candidate models, we adopt the method of Bayesian Information Criterion (BIC) to judge. The definition of BIC can be expressed as BIC=-2ln $L+k\cdot$ ln($n$), where $L$ is the maximum value of the likelihood function of the estimated model. $k$ and $n$ are the number of model parameters and data points, respectively. In fact, the value of BIC itself does not directly represent the absolute performance or predictive power of a model, but rather provides a relative metric for selecting among multiple models. BIC is a criterion to evaluate the best model fit among a finite set of models, and the model with the lowest BIC is preferred \citep{Neath2012}.

\begin{figure}
\centering
 \includegraphics [angle=0,scale=0.5] {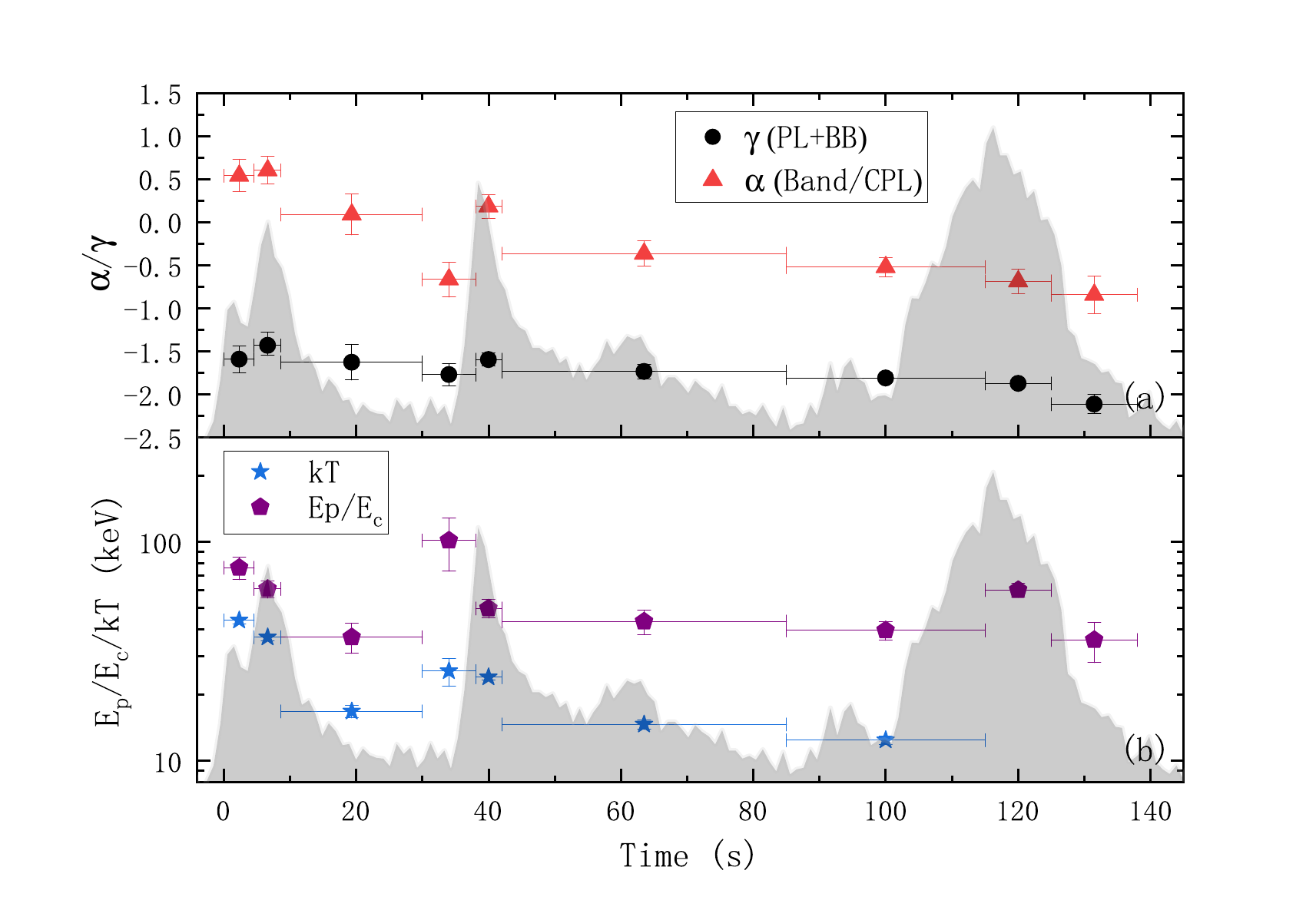}
 \caption{Temporal evolution of $\gamma$ (PL+BB) in panel (a) and $kT$ (BB or PL+BB) in panel (b). Low-energy index $\alpha$ from Band (or CPL) and $E_{\rm p}$ (or $E_{\rm c}$) are also shown in (a) and (b), respectively.}
 \label{fig:2}
\end{figure}

Then, we adopt the MCMC method to do the time-resolved spectral fitting, and the fitting results of time-resolved spectral analysis of GRB 161117A is shown in Table \ref{table1}\footnote{Note that the Band$+$BB and CPL$+$BB models yielded unconstrained fit parameters in all cases, so they are not listed in Table \ref{table1}}. It is found that the single BB model provides the best fit with well-converged parameters and the lowest BIC value for the first time interval. From the second to seventh time-bins, the PL$+$BB model is identified as the preferred model based on the BIC criterion. The non-thermal components, Band and CPL models are preferred for the 8th and 9th time-bins, respectively. It is easy to see that the jet components of GRB 161117A is undergoing the transition from purely thermal (BB) to a hybrid model (PL+BB), and finally to non-thermal (Band or CPL). It is worth noting that a different low-energy photon index $\alpha=0.4$ instead of $\alpha=1.0$ (e.g., $N(E) \propto E^{\alpha}$) is expected for a relativistic photosphere \citep{2010MNRAS.407.1033B}. Within this scenario, we also perform a comparative analysis using above blackbody model with $\alpha=0.4$, and found that the BIC value with $\alpha=0.4$ for first time interval is different from that of the standard BB model with $\alpha=1.0$ by only approximately 2. From the second to 9th time-bins, the fitting results are not changed. So that, we still adopt the standard BB model with $\alpha=1.0$ to do the spectral fitting and calculations in this work. On the other hand, we also present the temporal evolution of fitting parameters (e.g., $\alpha$, $\gamma$, $E_{\rm p}$, $E_{\rm c}$, and $kT$) for thermal and non-thermal emissions in Figure \ref{fig:2}, and find that the power-law index $\gamma$ is not significant evolution, but the $kT$ seems to be a hard-to-soft evolution during the first and second pulses. During the early phase of the third pulse, the temperature of thermal component is only 12.4 keV which is the lowest value by comparing with that of other thermal component.

\begin{figure}
\centering
 \includegraphics [angle=0,scale=0.5] {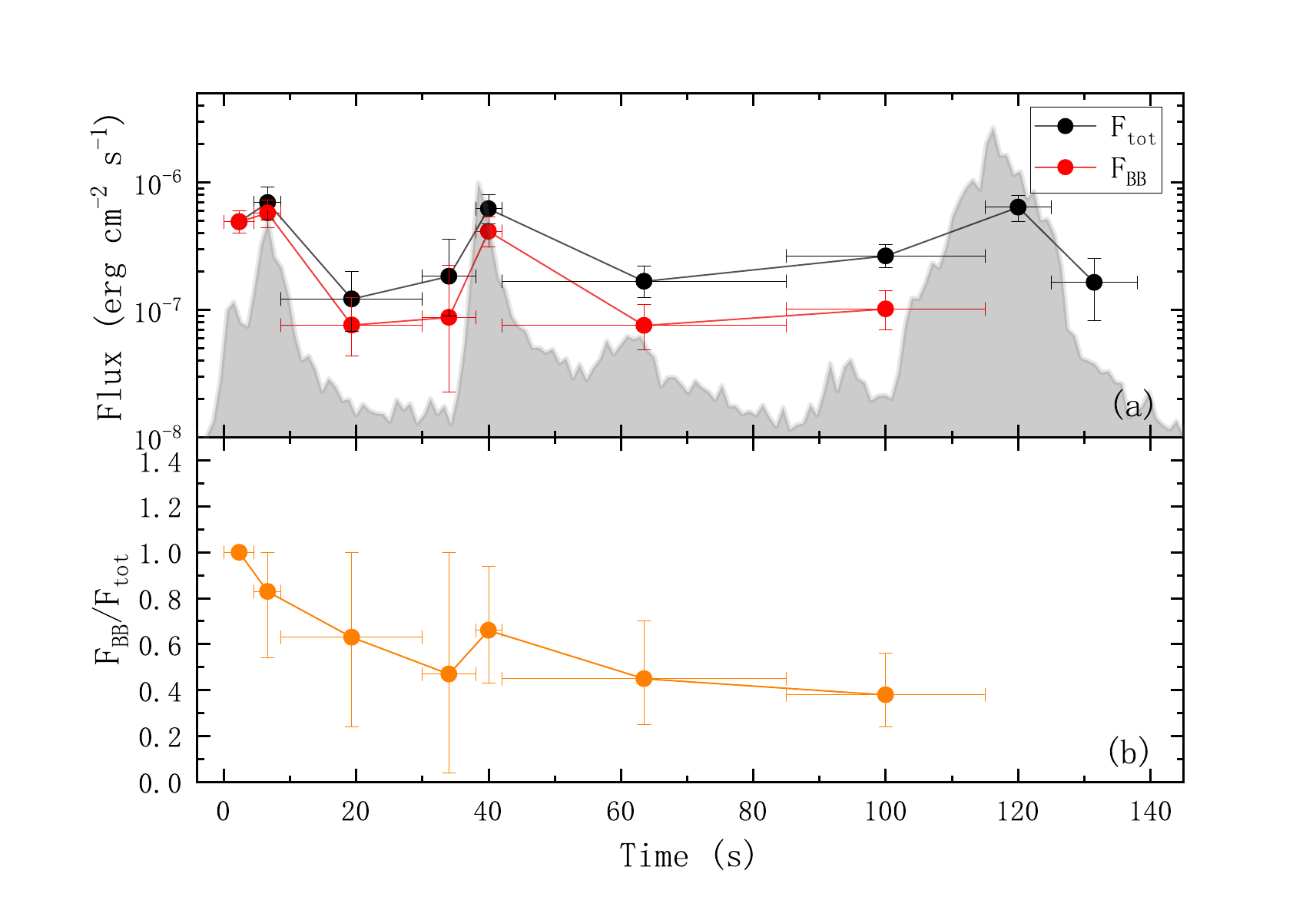}
 \caption{Temporal evolution of $F_{\rm BB}$ and $F_{\rm tot}$ in panel (a), as well as $F_{\rm BB}/F_{\rm tot}$ in panel (b).}
 \label{fig:3}
\end{figure}

Moreover, we also calculate the flux of BB emission ($F_{\rm BB}$), the total observed flux ($F_{\rm tot}$), and the ratio between $F_{\rm BB}$ and $F_{\rm tot}$ ($F_{\rm BB}/F_{\rm tot}$). Here, the flux calculated ranges from 10 keV to 1000 keV. In Figure \ref{fig:3}, we present the temporal evolution of $F_{\rm BB}$, $F_{\rm tot}$, and $F_{\rm BB}/F_{\rm tot}$. It is found that both $F_{\rm BB}$ and $F_{\rm tot}$ seem to be tracking with light curve of prompt emission. The calculated $F_{\rm BB}/F_{\rm tot}$ shows an overall hard-to-soft behavior with values ranging from 0.38 to 1.

\section{Derivation of the Physical Parameters within the Fireball Models}
In this Section, to understand more clearly of GRB 161117A physics, we use the significance quasi-thermal component in the prompt emission of GRB 161117A to derive relevant physical parameters, such as the Lorentz factor ($\Gamma_{\rm ph}$) and the radii of fireball \citep{2007ApJ...664L...1P}. We also calculate the magnetization factor ($\sigma_{0}$) and dimensionless entropy ($\eta$) at the central engine by adopting the hybrid jet of GRB 161117A.

\subsection{Lorentz Factor and Radius}
The time-resolved spectral of GRB 161117A shows that its jet composition is undergoing the transition from purely thermal (BB) to a hybrid model (PL+BB), and finally to non-thermal (Band or CPL). Following the method of \citet{2007ApJ...664L...1P}, one can estimate the $\Gamma_{\rm ph}$ based on the BB component in the time-resolved spectral of GRB 161117A, namely,
\begin{equation}
\Gamma_{\mathrm{ph}}=\left[1.06\times(1+z)^{2} D_{\mathrm{L}} \frac{Y \sigma_{\mathrm{T}} F_{\mathrm{tot}}}{2 m_{\mathrm{p}} c^{3} \Re}\right]^{1 / 4}
\end{equation}
where $D_{\rm L}$ is the luminosity distance, $m_{\rm p}$ is the proton mass, $\sigma_{\mathrm{T}}$ is the Thomson scattering cross-section. $\Re$ is defined as the ratio between the $F_{\rm BB}$ and the observed temperature $T$,
\begin{equation}
\Re \equiv (\frac{F_{\rm BB}}{\sigma T^{4}})^{1/2}
\end{equation}
Here, $\sigma$ is the Stefan's constant. $Y$ is the ratio between the total fireball energy and the energy emitted in the $\gamma$-ray, i.e.,
\begin{equation}
Y=\frac{E_{\gamma, \text { iso }}+E_{\mathrm{k}, \text { iso }}}{E_{\gamma, \text { iso }}}
\end{equation}
where $E_{\gamma, \text {iso}}$ and $E_{\mathrm{k}, \text {iso}}$ are the isotropic radiated energy in the prompt emission and the isotropic kinetic energy to power GRB afterglows, respectively. The isotropic radiated energy can be calculated by 
\begin{equation}
E_{\gamma, \text {iso}}=4\pi k^{*} D_{L}^{2}S_{\gamma}(1+z)^{-1}\sim 2.02 \times 10^{53}   \text {~erg},
\end{equation}
where $S_{\gamma}$ is the observed fluence, and $k^{*}$ is the k-correction factor from the observed band to $1–10^4$ keV in the burst rest frame \citep{2001AJ....121.2879B}. The isotropic kinetic energy $E_{\mathrm{k}, \text {iso}}$ can be estimated via the X-ray afterglow data \citep{2007ApJ...655..989Z,2014ApJ...785...74L}. For GRB 161117A, the XRT data are downloaded from the Swift data archive \footnote{https://www.swift.ac.uk/}. It is found that the X-ray light curve of GRB 161117A exhibits a plateau emission followed by a normal decay phase. By adopting a smoothly broken power-law to fit the data \citep{2007ApJ...670..565L}, one can obtain the decay slopes of $\alpha_1 \sim 0.31$, $\alpha_2 \sim 1.18$, and break time $t_{\rm b} \sim 7324$ s. Based on the relationship between decay index and spectral index, so called ``closure relation'', one can judge that it should be located in the spectral regime $\nu > \nu_c$ and interstellar medium (ISM) or Wind. Following the method from \citet{2007ApJ...655..989Z} and \citet{2014ApJ...785...74L}, the $E_{\mathrm{k}, \text {iso}}$ can be calculated as below in both ISM and wind
\begin{equation}
\begin{aligned}
    E_{\rm K,iso,52}= &\Big[  \frac{\nu F_{\nu}(\nu=10^{18}\rm Hz)}{5.2\times 10^{-14}{\rm~ erg~s^{-1}~cm^{-2}}}   \Big]^{4/(p+2)}\\ 
    &\times d_{z,28}^{8/(p+2)}(1+z)^{-1}t_d^{(3p-2/(p+2)}\\
    &\times (1 +Y^*)^{4/(p+2)}f_p^{-4/(p+2)}\epsilon_{B,-2}^{(2-p)/(p+2)}\\
    &\times\epsilon_{e,-1}^{4(1-p)/(p+2)}\nu_{18}^{2(p-2)/(p+2)}
\end{aligned}
\end{equation}
where $t_d$ is the time in the observer frame in days, $p$ is the electron’s spectral distribution index and calculated by $\alpha_2$ as $p \sim 2.24$. In our calculations, we adopt the typical values derived from observations and fixed the microphysics parameters of the shock $\epsilon_e=0.1$, $\epsilon_B=0.01$, and Compton parameter $Y^*=1$  \citep{2002ApJ...571..779P,2003ApJ...597..459Y}. $f_p$ is a function of the power law distribution electron spectral index $p$ \citep{2007ApJ...655..989Z}, 
\begin{equation}
    f_p \sim 6.73\Big(\frac{p-2}{p-1}  \Big)^{p - 1} (3.3\times 10^{-6})^{(p - 2.3)/2}
\end{equation}
If this is the case, we can calculate the $ E_{\mathrm{k}, \text {iso}} \sim 8.3 \times 10^{53}$ erg. One needs to note that the calculated $E_{\rm k, iso}$ is strongly dependent on the highly-uncertain physical parameters, such as $\epsilon_e, \epsilon_B$, and $Y^*$. Together with Eqs.(7) and (8), one can derive $Y\sim 5.1$. So that, the photosphere radius $R_{\rm ph}$ and the radius of central engine $R_{0}$ can be calculated as 
\begin{equation}
R_{\rm ph}=\left(L^{*} \sigma_{T} / 8 \pi \Gamma^{3}_{ph} m_{p} c^{3}\right)
\end{equation}
\begin{equation}
R_0 = \frac{4^{3/2}}{(1.48)^6(1.06)^4}\frac{D_L}{(1+z)^2}\left(\frac{F_\mathrm{BB}}{YF_\mathrm{tot}}\right)^{3/2}\mathcal{\Re}
\end{equation}
where $L^{*}=4 \pi Y D_{\rm L}^{2}F_{\rm tot}$ is the luminosity. The calculated results are shown in Table \ref{table2}.

\begin{figure}
\centering
 \includegraphics [angle=0,scale=0.5] {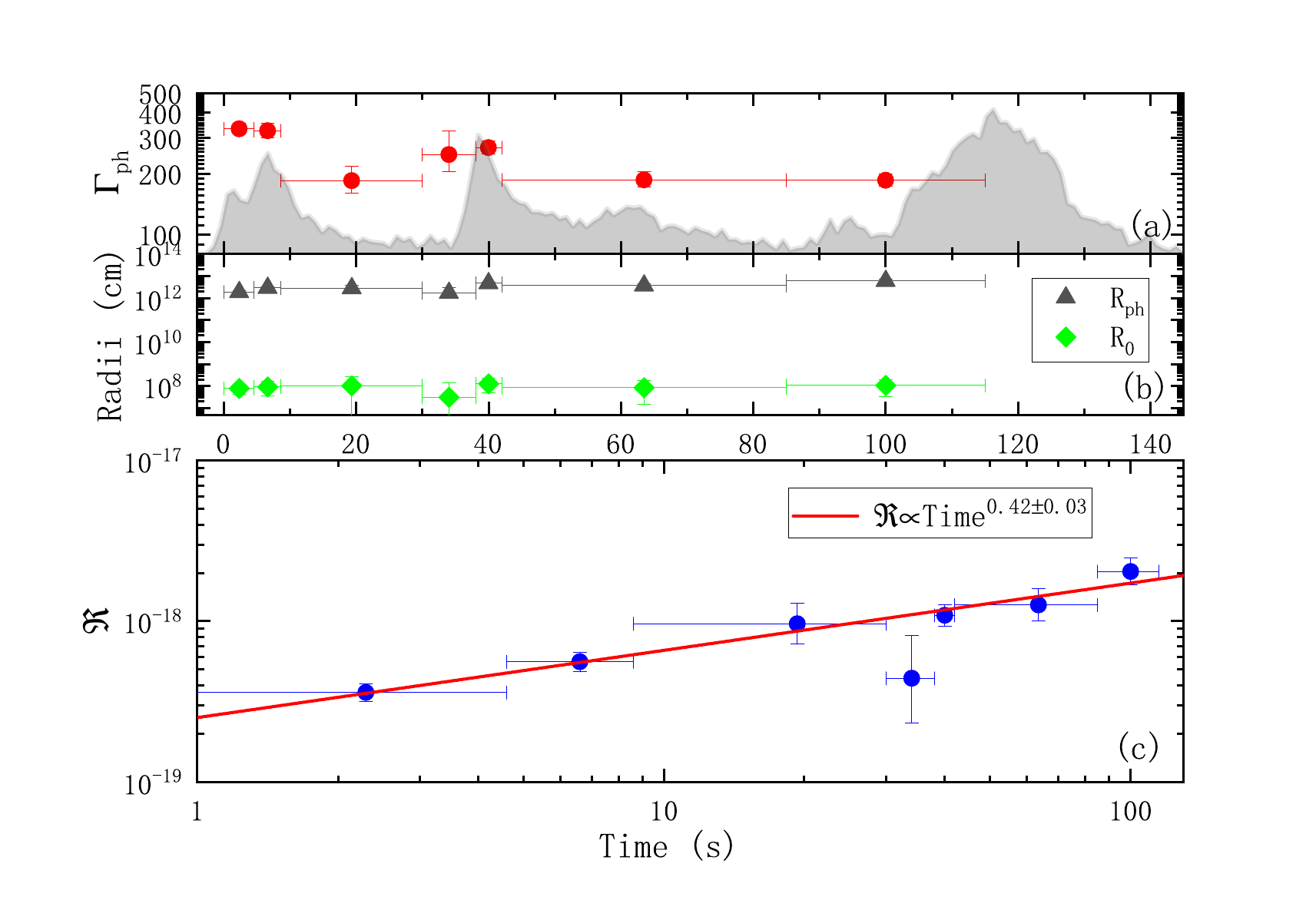}
 \caption{Temporal evolution of $\Gamma_{\rm ph}$ in panel (a), and radius ($R_{0}$ and $R_{\rm ph}$) in panel (b), as well as the derived $\Re$ in panel (c).}
 \label{fig:4}
\end{figure}

Figure \ref{fig:4} shows the temporal evolution of $\Gamma_{\rm ph}$, $R_{\rm ph}$, $R_{0}$, and $\Re$. We find that the evolution of $\Gamma_{\rm ph}$ seems to be tracking with the light curve, and the maximum value can be reached to 333 at the initial of the light curve. Then, it is gradually going down to 185 and keep almost constant. The $R_{\rm ph}$ is a little bit to increase within the range of $[1.7 \times 10^{12} - 6 \times 10^{12}]$ cm, and $R_0$ is almost constant around $\sim$ $10^{8}$ cm. On the other hand, it is found that the temporal behavior of the $\Re$ exhibits a power-law rising with index of approximately 0.42, and it suggests that the absence of significant energy dissipation is below the photosphere radius \citep{2007ApJ...664L...1P}.

\subsection{Magnetization Factor and Dimensionless Entropy}
\citet{2015ApJ...801...2} introduced a hybrid relativistic outflow model for GRBs which includes a hot fireball component (defined by dimensionless entropy $\eta$) and a Poynting-flux component (defined by magnetization $\sigma_0$). One interesting question is whether the observed non-thermal component in the hybrid spectra of GRB 161117A originates from the internal shock of the fireball or an additional cold Poynting-flux component? In this Section, we infer both $\sigma_0$ and $\eta$ within the assumption of a hybrid jet of GRB 161117A. Specifically, the central engine magnetization factor $\sigma_{0}$ is defined as $\sigma_{0}=L_{\rm c}/L_{\rm h}$, where the $L_{\rm h}$ and $L_{\rm c}$ are the luminosity of hot fireball component and cold Poynting-flux component, respectively. The dimensionless entropy $\eta$ can be defined as $\eta=L_{\rm h}/\dot{M}c^2$, where $\dot{M}$ is the accretion rate. Different values of ($\eta$, $\sigma_0$) pairs correspond to evolution of photosphere emission proprieties \citep{2015ApJ...801...2}. For example, (1) ${\bf \eta \gg}$ 1, $\sigma_{0}\ll$1: the photosphere emission is primarily driven by a pure fireball component; (2) ${\bf \eta \sim}$ 1, ${(1+\sigma_{0})\gg}$ 1: the outflow is dominated by the Poynting-flux.
By adopting the judgment criteria proposed by \citet{2015ApJ...801...2}, based on the observed flux ($F_{\rm BB}$ and $F_{\rm tot}$) and temperature ($kT$), one can infer the parameters of the central engine, such as $1+\sigma_0$, $\eta$, $\Gamma_{\rm ph}$, and the photosphere radius $R_{\rm ph}$. It is also so-called "top-down" method proposed by \citep{2015ApJ...801...2}. 

However, the inferred parameters of the central engine strongly depend on the chosen initial radius ($R_{0}$) of the central engine. By adopting the result of Eq. (12), the derived parameters of central engine for GRB 161117A within a hybrid jet model are shown in Table \ref{table2}. There are six distinct regimes to characterize the photosphere properties in the hybrid system, and they are applicable to outflows in scenarios with both sub-photospheric magnetic dissipation and non-sub-photospheric magnetic dissipation. Regime I: $\eta>(1+\sigma)^{1 / 2}$ and $r_{\rm ph}<r_{\rm ra}$; Regime II: $\eta>(1+\sigma)^{1 / 2}$ and $r_{\rm ra}<r_{\rm ph}<r_{\rm c}$; Regime III: $\eta>(1+\sigma)^{1 / 2}$ and $r_{\rm ph}>r_{\rm c}$; Regime VI: $\eta<(1+\sigma)^{1 / 2}$ and $r_{\rm ph}<r_{\rm ra}$; Regime V: $\eta<(1+\sigma)^{1 / 2}$ and $r_{\rm ra}<r_{\rm ph}<r_{\rm c}$; Regime VI: $\eta<(1+\sigma)^{1 / 2}$ and $r_{\rm ph}>r_{\rm c}$. Similarly, due to the degeneracy in the hybrid problem, the central engine parameters cannot be inferred for $r_{\rm ph} < r_{\rm ra}$ (corresponding to Regimes I and IV). Thus, our analysis will concentrate on the case for $r_{\rm ph} > r_{\rm ra}$ (i.e., Regimes II, III, V, and VI). Here, $r_{\rm ra}$ is the radius of rapid acceleration, and $r_{\rm c}$ is the coasting radius. The calculated details are shown in Appendix, also can refer to \citep{2015ApJ...801...2}.

\begin{figure}
\centering
 \includegraphics [angle=0,scale=0.5] {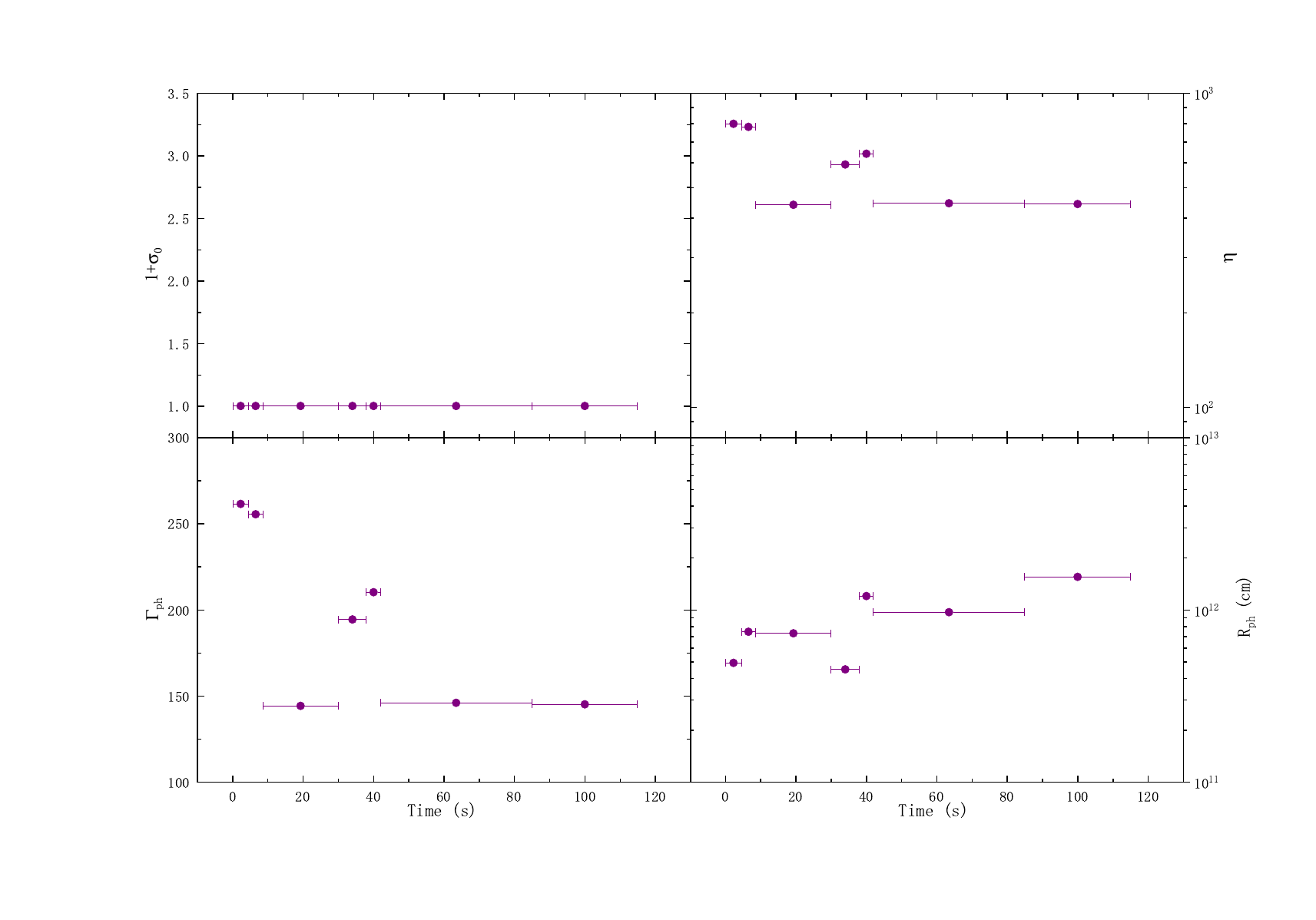}
 \caption{Temporal evolution of $1+\sigma_{0}$, $\eta$, $\Gamma_{\rm ph}$ and $R_{\rm ph}$ in the hybrid jet model of GRB 161117A.}
 \label{fig:5}
\end{figure}

Figure \ref{fig:5} shows the temporal evolution of $1+\sigma_0$, $\eta$, $R_{\rm ph}$ and $\Gamma_{\rm ph}$. It is found that the evolution behaviors of $\Gamma_{\rm ph}$ and $R_{\rm ph}$ within the hybrid jet model is similar to that within a pure fireball model \citep{2007ApJ...664L...1P}. Moreover, we also find the values $1+\sigma_0$ $\sim$ 1 and $\eta \gg$ 1 for all time slices including a thermal component (from the 1st to 7th), and it suggests that the outflow of those time bins with thermal component are at least matter-dominated rather than that of Poynting-flux. The non-thermal component (e.g., PL component) mainly arise from particle acceleration via internal shocks rather than magnetized dissipation.

\section{Constraint on Magnetization Parameter with Purely Non-thermal Spectra}
Based on the results in Section 3.2, it is clear to see that the spectra with thermal component (e.g., pure thermal and hybrid) are dominated by a purely fireball without any Poynting-flux component. Another interesting question is whether the purely non-thermal spectra during the late phase of prompt emission (e.g., 8th and 9th time intervals) are dominated by Poynting-flux-outflows. \citet{2009ApJ...700L..65Z} proposed a method to constraint the magnetization factor at the photosphere radius ($\sigma_{\rm ph}$) by assuming a pseudo and undetected blackbody spectrum. We apply the method proposed by \citet{2009ApJ...700L..65Z} to constrain the $\sigma_{\rm ph}$ of two purely non-thermal spectral during the late prompt emission of GRB 161117A. More details of this method can refer to \citet{2009ApJ...700L..65Z} and \citet{2024MNRAS.529L..67D}.

\begin{figure}
\centering
 \includegraphics [angle=0,scale=0.3] {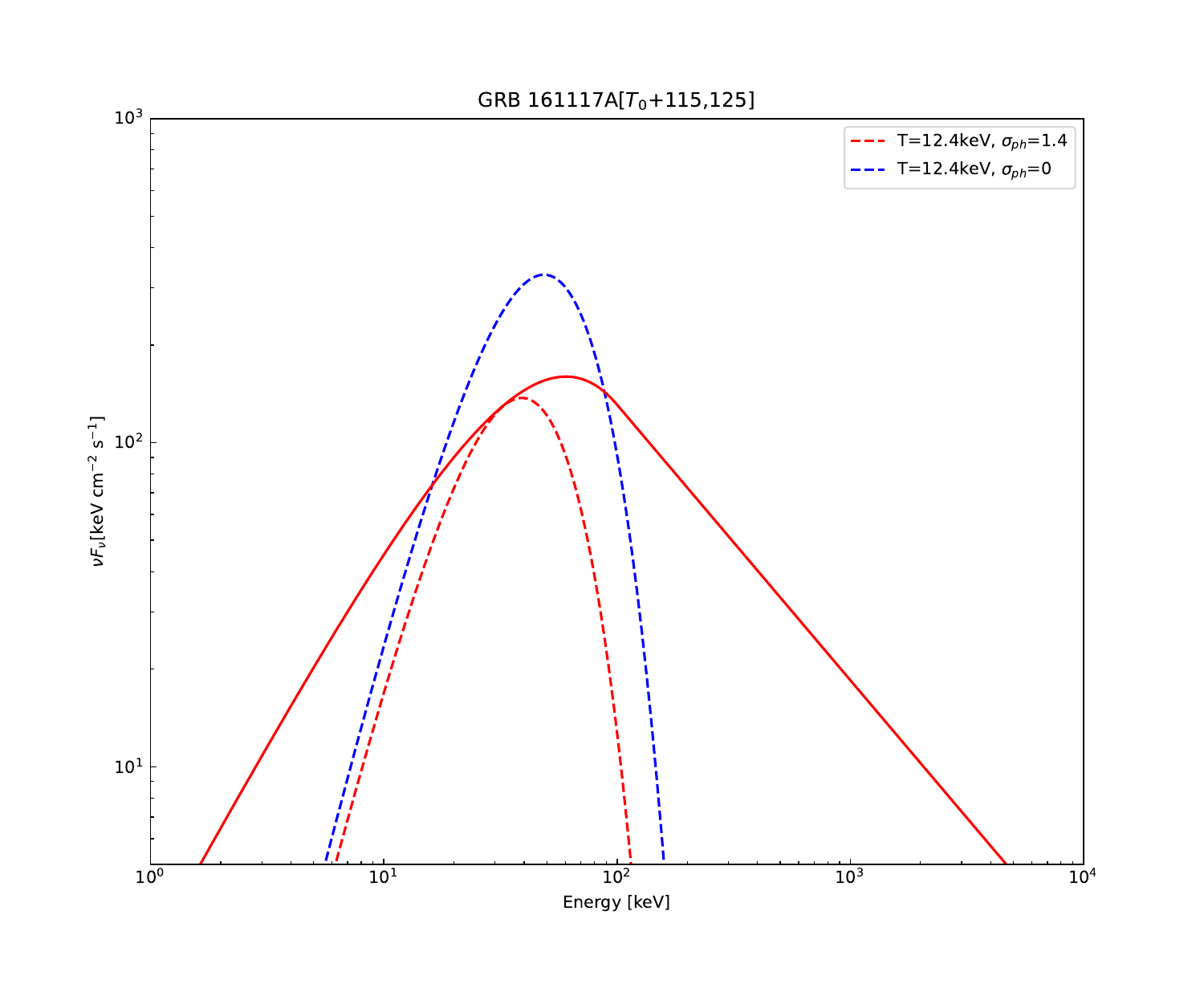}
 \includegraphics [angle=0,scale=0.3] {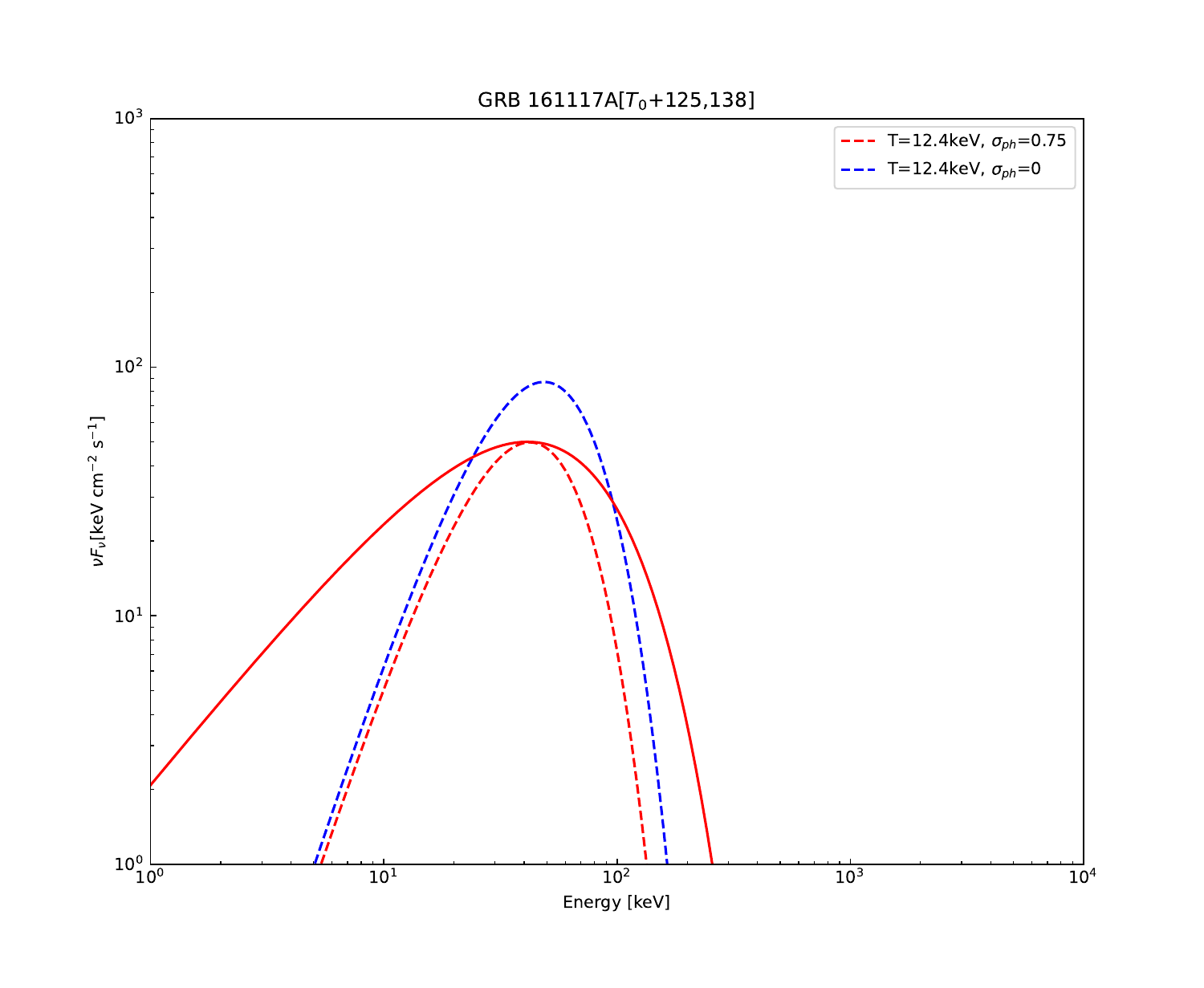}
 \caption{Left: The observed Band spectrum of the 8th time slice and the predicted lower limits of the photosphere pseudo-thermal spectra (The red and blue dashed lines represent magnetization factors equal to 1.4 and 0, respectively.) for temperature fixed on 12.4 keV. Right: Similar to the left panel, but for the CPL spectrum of the 9th time slice (The red and blue dashed lines represent magnetization factors equal to 0.75 and 0, respectively).}
 \label{fig:6}
\end{figure}

Observationally, the non-thermal spectral energy distribution (SED) of the last two time-intervals for GRB 161117A can be described by Band and CPL model, respectively. Furthermore, we assume that the temperature $kT=12.4$ keV measured in the 7th time-interval is adopted to calculate the thermal luminosity, and a pseudo-blackbody spectrum is produced by the photosphere. Following the method described in \citet{2009ApJ...700L..65Z} and \citet{2024MNRAS.529L..67D}, one can calculate the lower limit of the expected photosphere spectrum based on the internal shock model within a baryon-dominated outflow \footnote{Here, we assume that the radiative efficiency is 100\%, and the wind luminosity is given by $L_{\rm w} = L_{\gamma}$, where $L_{\gamma}$ is the gamma-ray luminosity.} (see Figure \ref{fig:6}). Then, we compare it with the observational data, and we find that the pseudo thermal emission is higher than that of observed non-thermal emission for the magnetization parameter $\sigma_{\rm ph}=0$. If this is the case, in order to suppress the bright thermal emission, one can infer a lower limit of the magnetization parameter ($\sigma_{\rm ph}=L_{\mathrm{p}}/L_{\mathrm{b}}$), which is defined as the ratio between the Poynting flux ($L_{\mathrm{p}}$) and the baryonic flux ($L_{\mathrm{b}}$). The wind luminosity can be rewritten as $L_{\mathrm{w}}=L_{\mathrm{p}}+L_{\mathrm{b}}=(1+\sigma_{\rm ph})L_{\mathrm{b}}$ \citep{2002ApJ...581.1236Z}. According to the Eqs. (1) (2) (3) in \citet{2024MNRAS.529L..67D}, the $L_{\rm w}$ can be replaced with $L_{\rm w}/(1+\sigma_{\rm ph})$. So, one can estimate a minimum value of $\sigma_{\rm ph}$ which can be used to suppress the expected thermal component from photosphere emission. 

In Figure \ref{fig:6}, we plot the predicted lower limits of the photosphere pseudo-thermal spectra for last two time-interval of GRB 161117A prompt emission, and the red dotted line is the thermal emission that makes the photosphere emission undetected. It is found that the required values of $\sigma_{\rm ph}$ for the last two time-interval at least are 1.4 and 0.75, respectively. Here, we fixed the temperature as 12.4 keV which is the blackbody temperature observed with minimum temperature detected in the 7th time bin. Based on the observed hard-to-soft evolution trend of temperature within individual pulses, 12.4 keV is selected as the maximum temperature of the non-thermal spectrum to estimate the magnetization parameter. Such low $\sigma_{\rm ph}$ values suggest that at least the internal shock model is a viable mechanism to interpret GRB 161117A. However, we cannot completely rule out Poynting-flux outflows, as the $\sigma_{\rm ph}$ values represent only a lower limit.

\section{Possible physical interpretation: NDAF Model}
As discussed above, no evidence of Poynting flux in the time-resolved spectra of GRB 161117A, suggests that the BZ mechanism does not seem to apply for GRB 161117A. \citet{2024ApJ...975..225L} proposed that the thermal component in the GRB spectra may originate from $\nu \bar{\nu}$ annihilation. In this Section, we will attempt to adopt the $\nu \bar{\nu}$ annihilation mechanism of the NDAF model to explain the thermal emission of GRB 161117A.

\citet{2013ApJS..207...23X} calculated one-dimensional global solutions of NDAF, and the annihilation luminosity and height can be approximately written as
\begin{equation}
\log L_{\nu \bar{\nu}}\left(\mathrm{erg} \mathrm{~s}^{-1}\right) =52.98+3.88 a_{*}-1.55 \log m_{\mathrm{BH}} 5.0 \log \dot{m}
\end{equation}
\begin{equation}
\log h=2.15-0.30 a_{*}-0.53 \log m_{\mathrm{BH}}+0.35 \log \dot{m},
\end{equation}
where $m_{\rm BH}=M_{\rm BH}/M_{\odot}$ and $\dot{m}=\dot{M}/M_{\odot}$ $\rm s^{-1}$ are the dimensionless BH mass and accretion rate, respectively. $0\le a_{*}\le1$ is the dimensionless BH spin parameter, $H=h\cdot r_{g}$ is the annihilation height, which is defined as the region where 99.9\% of the annihilation luminosity is included, and $r_{g}=2GM_{BH}/c^2$ is the Schwarzschild radius. 

From the observational point of view, the mean jet power of GRBs can be estimated as \citep{2011ApJ...739...47F,2015ApJ...806...58L}
\begin{equation}
P_\mathrm{j}\simeq\frac{    (E_{\gamma,\mathrm{iso}}+E_\mathrm{k,iso}) (1+z)\theta_\mathrm{j}^2}{2T_{90}}.
\end{equation}
Here, $\theta_{j}$ is the jet-opening angle which can be derived from observational data \citep{1999ApJ...525..737R,1999ApJ...519L..17S,2001ApJ...562L..55F},
\begin{equation}
\begin{aligned}
\theta_{j}= & 0.124\left(\frac{t_{j}}{1 \text { day }}\right)^{3 / 8} \times\left(\frac{1+z}{2}\right)^{-3 / 8}\left(\frac{E_{\mathrm{k}, \mathrm{iso}}}{10^{52} \mathrm{erg}}\right)^{-1 / 8}\left(\frac{n^{*}}{1 \mathrm{~cm}^{-3}}\right)^{1 / 8}
\end{aligned}
\end{equation}
where $t_j$ is the jet-break time, and $n^{*}$ is the density of the constant ambient medium. In our calculations, we adopt the observed time of the last data point ($\sim$ 15.8 days) as the lower limit of jet-break time ($t_j$) and fix the ambient medium density as $n=1\rm~cm^{-3}$. One can estimate the lower limit of $\theta_{j}\sim 10.5^{\circ}$. Therefore, by adopting the $E_{\gamma, \text {iso}}$ = $2.02 \times 10^{53}$ erg, $E_{\mathrm{k}, \text {iso}} \sim 8.3 \times 10^{53}$ erg, $\theta_{j} \sim 0.18$ rad, and $T_{90}\sim 124$ s, one can derive the lower limit of jet power $P_{\rm j}\sim 3.57\times 10^{50}$ erg $\rm s^{-1}$. Based on Eqs. (13) (14) and (15), namely, $P_{\rm j}= L_{\nu \bar{\nu}}$ for typical ranges of the BH mass (3 $\le m_{\rm BH}\le$ 10), one can obtain the annihilation height to compare with the central engine radius $R_{0}$, here we adopt the average ${R}_{0}\sim 9.1\times 10^{7}$ cm. 

\begin{figure}
\centering
 \includegraphics [angle=0,scale=0.3] {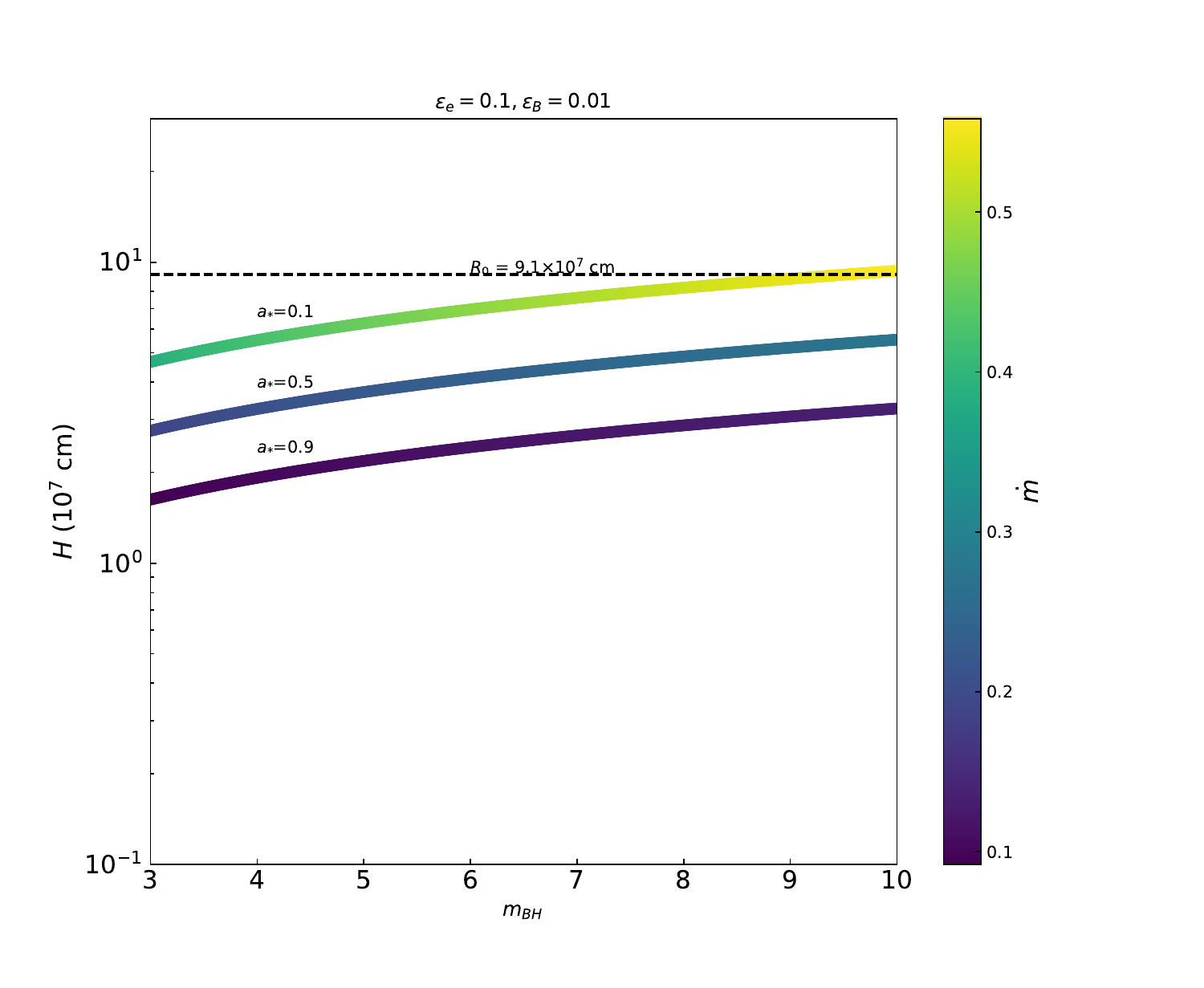}
  \includegraphics [angle=0,scale=0.3] {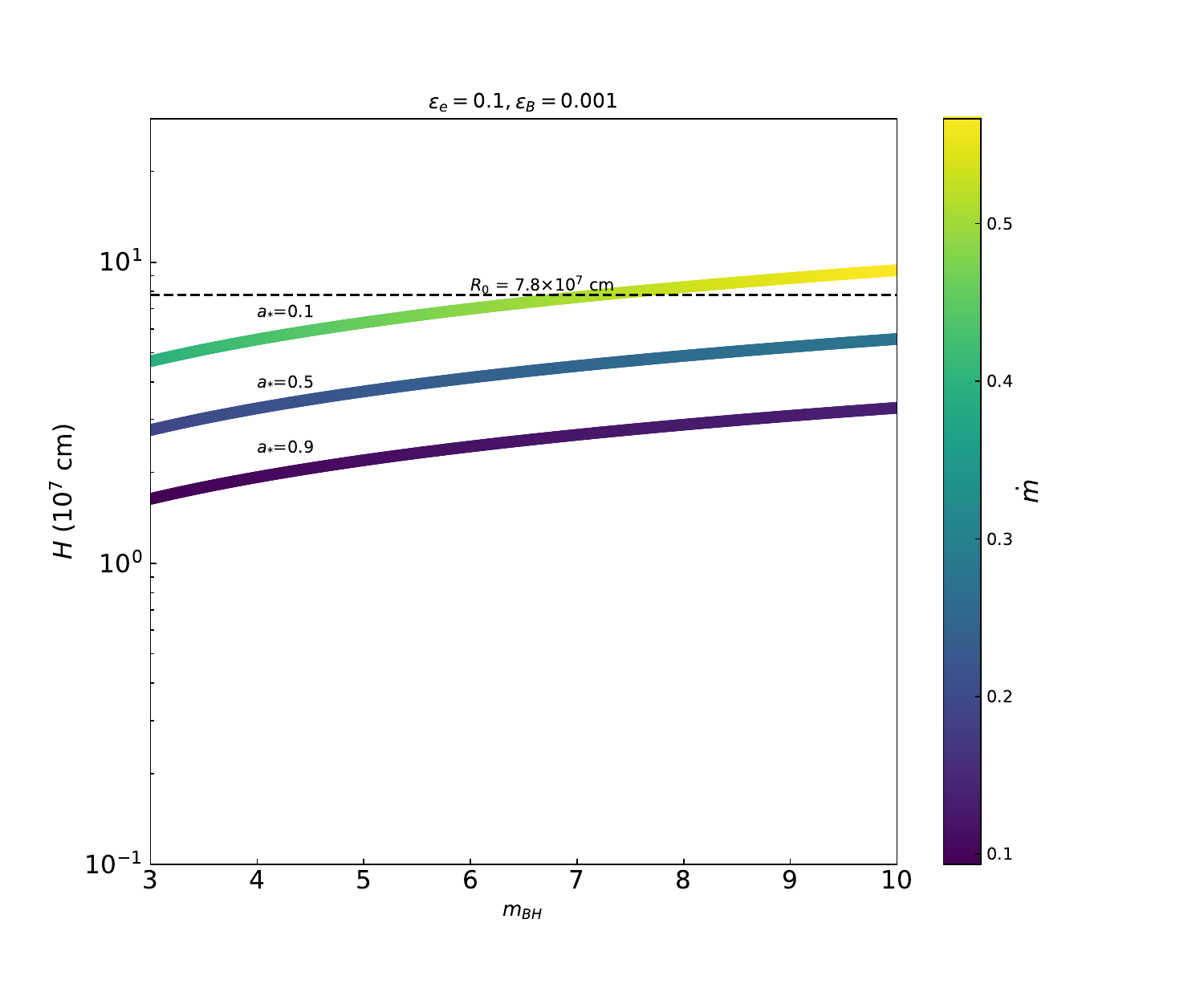}
   \includegraphics [angle=0,scale=0.3] {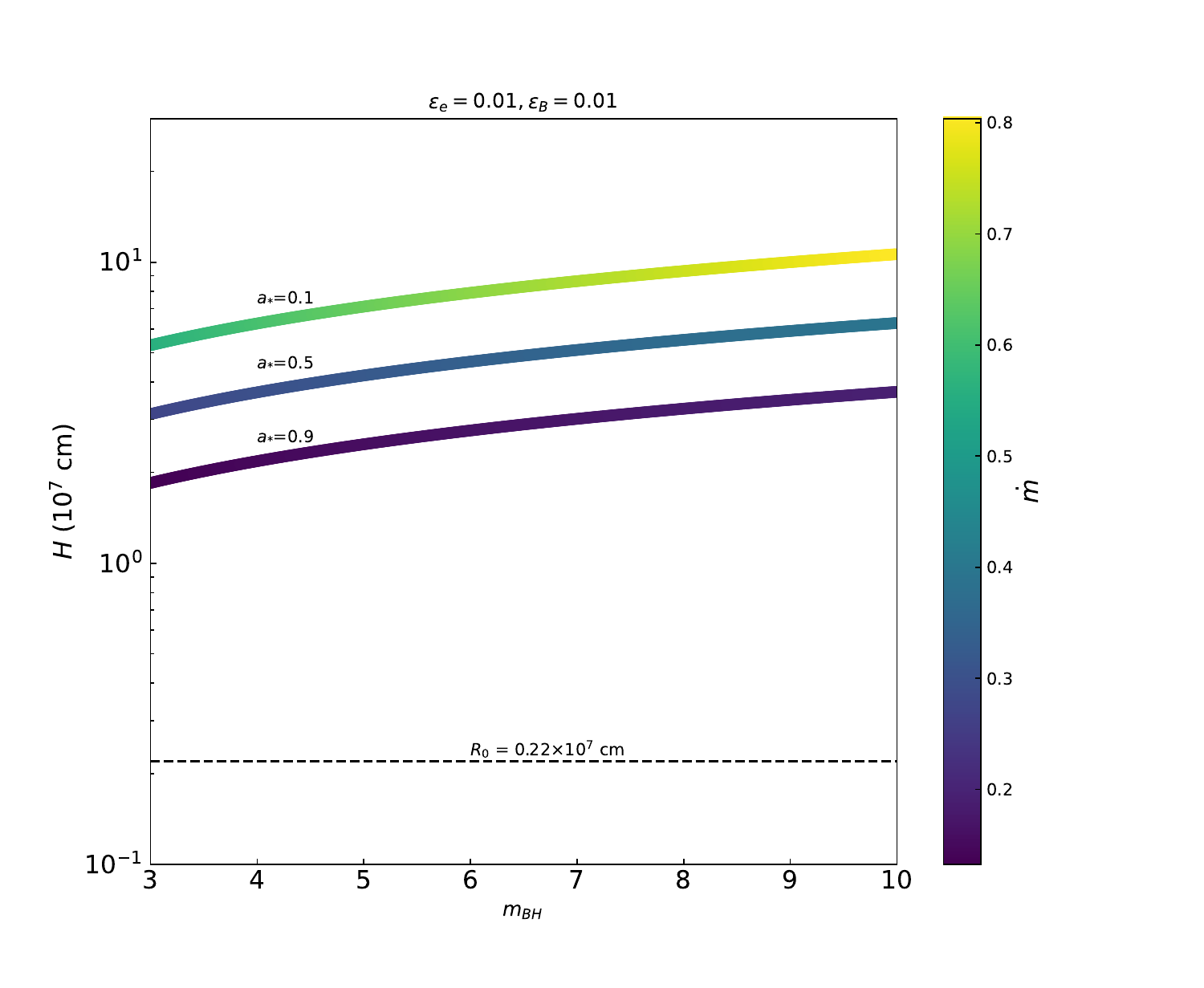}
    \includegraphics [angle=0,scale=0.3] {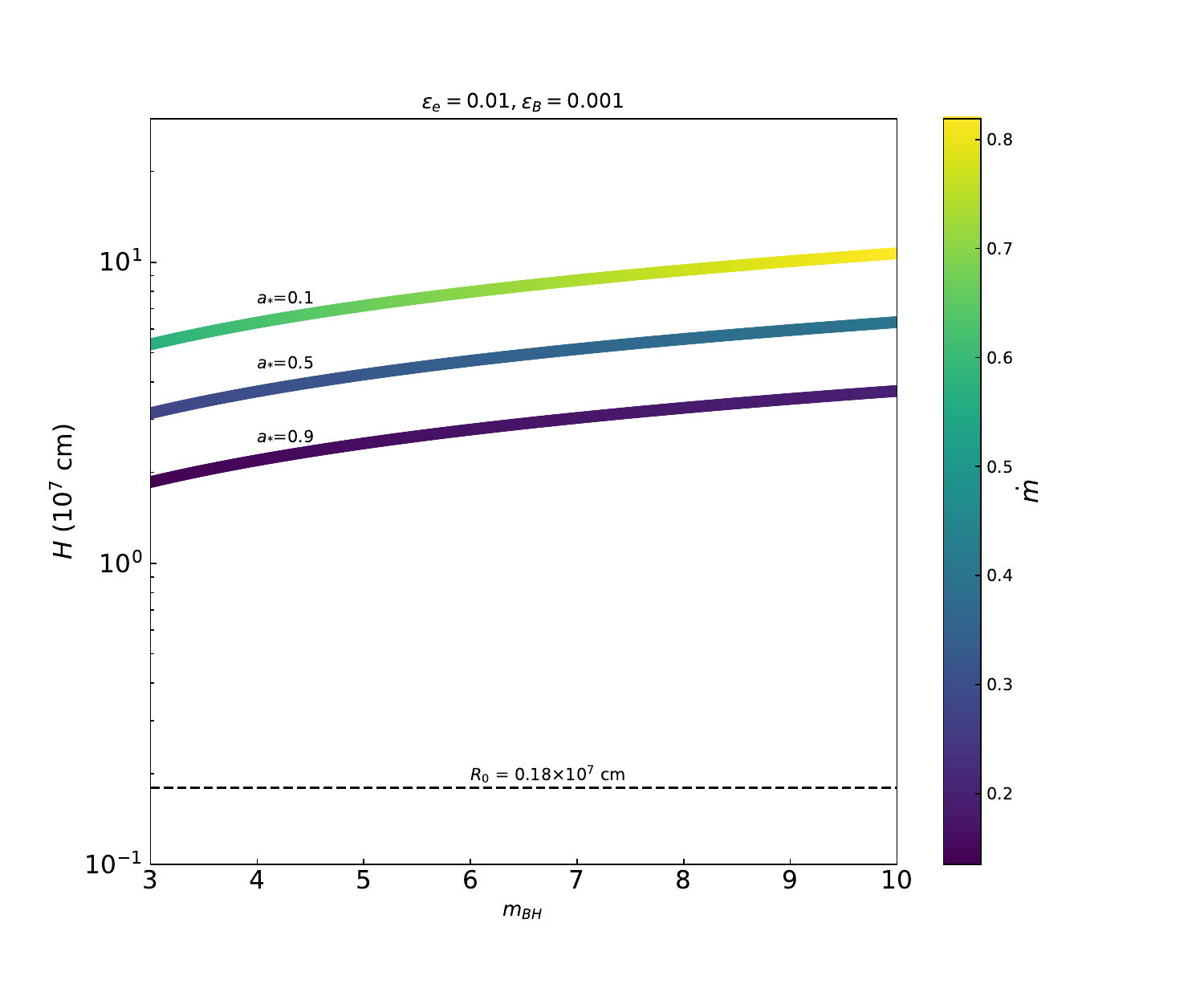}
 \caption{Neutrino annihilation height $H$ as a function of BH mass and the accretion rates in different BH spin parameter $a_{*}$. The multicolor regions denote the range of the dimensionless accretion rates, and the black dashed lines correspond to the central engine radius of the GRB fireball.}
 \label{fig:7}
\end{figure}

Figure \ref{fig:7} shows a three-dimensional plot regarding $H$, $m_{\rm BH}$, and $\dot{m}$. It is found that the ${R}_{0}$ is almost covered by the reasonable region of annihilation height $H$, and the $\dot{m}$ ranges of [0.09 - 0.56] for $a_{*}\in [0.1,0.9]$. It suggests that the NDAF model could be used to explain the thermal emission in GRB 161117A, and the $\nu \bar{\nu}$ annihilation mechanism can account for the jet of GRB 161117A. If this is the case, the thermal component and non-thermal emission of GRB 161117A are possible to come from the photosphere emission and the internal shocks, respectively. Moreover, the calculated results depend on the highly-uncertain parameters, such as $\epsilon_e, \epsilon_B$, and an external density $n^{*}$. So that, we also adopt different values of $\epsilon_e$ = 0.1 and 0.01, and $\epsilon_B$ = 0.01,0.001 to calculate the annihilation height $H$, and the comparison diagrams are also shown in Figure \ref{fig:7}. It is found that when $\epsilon_e$ = 0.1, the average ${R}_{0}$ nearly covers the entire range of annihilation height $H$ regardless of whether $\epsilon_B$ = 0.01 or 0.001, indicating that the results are insensitive to the choice of $\epsilon_B$. However, when $\epsilon_e$  = 0.01, the annihilation height significantly exceeds  $R_{0}$, making the results highly sensitive to the value of $\epsilon_e$.

\section{Conclusion and Discussion}
GRB 161117A exhibits three overlapping broad pulses with duration $T_{90}\sim 124$ s, and the measured redshift is $z=1.549$. In this work, we adopt the data observed with GBM on board the Fermi mission during the prompt emission phase, and perform a detailed time-resolved spectral analysis of GRB 161117A using Bayesian and MCMC methods. The light curve is divided into nine time intervals to do the spectral fitting, applying six different spectral models (e.g., Band, CPL, BB, PL$+$BB, Band$+$BB, and CPL$+$BB). Then, we compare with those models by calculating the values of BIC, and select the favourite model with the lowest of BIC. By analyzing the data and deriving the physical parameters, we find the following interesting results: 
\begin{itemize}	
\item[$\bullet$] A single BB model can adequately fit the spectra of the first time-interval of GRB 161117A prompt emission, while the best model of spectral fitting is the PL$+$BB model from the second to seventh time-interval. Finally, the Band and CPL models are the best one to do the spectral fitting for the last eighth and ninth time-interval, respectively. Those results suggest that an evolutionary trend of spectra do exist from thermal to hybrid and finally to non-thermal emission in the prompt emission of GRB 161117A. 
\item[$\bullet$] By adopting the quasi-thermal component in the prompt emission of GRB 161117A to derive the physical parameters, it is found that the temporal evolution of the Lorentz factor ($\Gamma_{\rm ph}$) seems to be tracking with the light curve ranging from 185 to 333, and the $R_0$ remains to be a constant around $\sim$ $10^{8}$ cm. On the other hand, the calculated $1+\sigma_0$ $\sim$ 1 and $\eta \gg$ 1 of all thermal component in the time-interval (1st to 7th), suggest that the outflow of those time-interval are matter-dominated rather than Poynting-flux-dominated, and the non-thermal component (e.g., PL component) mainly arise from particle acceleration via internal shocks rather than magnetized dissipation.
\item[$\bullet$] By simulating pseudo-thermal spectra, we estimate the magnetization parameter at photosphere radius $\sigma_{\rm ph}$ of the purely non-thermal spectra during the late prompt emission phase of GRB 161117A, and find that the lower limits of the $\sigma_{\rm ph}$ for the last two time-interval (e.g,. 8th and 9th) are 1.4 and 0.75, respectively. It suggests that at least the internal shock model is a viable mechanism to interpret GRB 161117A at late time. However, we cannot fully rule out the possibility of Poynting-flux outflows due to the rough lower limit of the $\sigma_{\rm ph}$ values.
\end{itemize}

One possible interpretation of such spectral transition from thermal to hybrid and finally to non-thermal emission in the prompt emission of GRB 161117A is from $\nu \bar{\nu}$ annihilation mechanism of NDAF model \citep{2024ApJ...975..225L}. By invoking the method from \citet{2024ApJ...975..225L}, it is found that the initial size of the flow $R_{0}$ is almost covered by the reasonable region of annihilation height $H$, and suggests the $\nu \bar{\nu}$ annihilation mechanism can account for the jet of GRB 161117A, while the thermal component and non-thermal emission of GRB 161117A are possible to come from the photosphere emission and internal shocks, respectively. If the central engine of GRB 161117A is indeed a black hole-accretion disk system, the gravitational energy released from the accretion disk via the $\nu \bar{\nu}$ annihilation mechanism can drive the jet. Within this scenario, the observed thermal and non-thermal emissions in the hybrid jet (e.g., from the second to 7th time-interval) should be contributed from photosphere emission and internal shocks, respectively, and it does not require any Poynting-flux in the jet composition during this phase. However, we only provide lower limits on constraining the magnetization factor ($\sigma_{\rm ph}$) which is not high enough at photosphere in last two time-intervals. If the real $\sigma_{\rm ph}$ is much larger than the lower limit, the BZ mechanism can extract the black hole’s rotational energy via magnetic fields (e.g., Poynting-flux) to produce the observed non-thermal component in the relativistic jet. It is worth noting that the calculated $E_{\rm k, iso}$ and $P_{\rm j}$ are dependent on the highly-uncertain physical parameters, such as $\epsilon_e, \epsilon_B$, $Y^*$, an external density $n^{*}$.

On the other hand, it is possible that there are two components in all intervals but that the detectability of the weakest component becomes difficult when the other component is strongly dominated. For the first time interval of GRB 161117A, the best-fit model is a pure blackbody radiation, and subtle fluctuations in sub-photospheric energy dissipation could be the possible origin of the variability light curve. Moreover, adding a non-thermal component (PL or Band) to this time bin yielded no significant fit improvement ($BIC_{\rm BB}< BIC_{\rm PL/Band+BB}$), suggesting that the non-thermal emission may be below the detection limit. For time bins 8–9 with only non-thermal spectra (Band/CPL), adding a BB component failed statistical validation (unconverged posterior distribution or larger BIC), indicating that the thermal component is either absent to be sub-detectable, or suppressed by the dominant non-thermal emission. Notably, such sub-component flux is too low to affect analysis results. Furthermore, in the early phase of GRB 161117A, pure blackbody emission coexists with slight variability (internal shock component undetected), the reason can be attributed to two ways. One is from the temperature evolution of the thermal component to result in the slight variability of light curve, and it is quite similar to the case of GRB 131014A \citep{2015ApJ...814...10G}. The other one is that the sub-photospheric energy dissipation fluctuations (relativistic turbulence in the emission region or jet-envelope interaction) can also result in such slight variability of light curve \citep{2015PhR...561....1K}). In any case, both possibilities can be used to explain the minor variability in the light curve of GRB 161117A without requiring an internal shock component.

Based on the spectral evolution of GRB 161117A, namely, a transition from thermal to hybrid, and finally to non-thermal emission, the physical picture can be described as follows: Initially, the $\nu \bar{\nu}$ annihilation is dominated with extremely high accretion rate of black hole to produce the purely thermal emission (e.g., first time-interval) when the photons escape the system at the photosphere radius; Then, the non-thermal component (e.g., from second to 7th time-interval) can be produced by internal shock in the hybrid jet when the photons reach to the internal shock radius; Finally, the $\nu \bar{\nu}$ annihilation is decreased with time because of decreasing accretion rate of black hole at late time. Meanwhile, the magnetic field of the black hole becomes gradually stronger and stronger due to cumulated time, and the BZ mechanism may be dominated to produce purely non-thermal emission at late time. It is worth noting that above physical process is only one possible interpretation for the prompt emission of GRB 161117A, and we can not rule out other explanations for this case. Multi-messenger and multi-band observations can be used to explore the jet and radiation physical process of GRB 161117A-like event in the future.

\begin{acknowledgements}
We acknowledge the use of the public data from the Fermi/GBM and Swift/XRT data archive. This work is supported by the Guangxi Science Foundation (grant nos 2023GXNSFDA026007 and 2025GXNSFDA02850010), the National Natural Science Foundation of China (grant Nos. 12494574 and 12133003), the Program of Bagui Scholars Program (LHJ), and the Guangxi Talent Program (“Highland of Innovation Talents”).

\end{acknowledgements}

\begin{table*} \centering
\renewcommand\tabcolsep{2.2pt}
\renewcommand\arraystretch{1.3}
\caption{The spectral analysis results of GRB 161117A.}
\begin{threeparttable} 
\begin{tabular}{ccccccccccc} 
\hline\hline
$t_{1}-t_{2}$(s)&$\rm S/N$&$\rm Model$&$\alpha(\gamma)^a$&$E_{\rm p}/E_{\rm c}/kT \rm ^{b}$&$\beta$&$F_{\rm tot}^{~~c}$&$F_{\rm BB}^{~~c}$&BIC&\rm favourite model \\
\hline
0.0 	$\sim$	4.6 	&	22 	&	CPL		&$	0.54 	_{	-0.19 	}^{+	0.19 	}$&$	76.2 	_{	-9.0 	}^{+	8.9 	}$&						&	$	0.57 	^{+	0.43 	}_{-	0.25 	}$&						&		3061.9 	&		\\
$\quad$			&		&	Band		&$	0.57 	_{	-0.21 	}^{+	0.22 	}$&$	187.0 	_{	-12.0 	}^{+	12.4 	}$&$	-3.38 	_{	-0.42 	}^{+	0.43 	}$&	$	0.60 	^{+	0.23 	}_{-	0.16 	}$&						&		3081.5 	&		\\
$\quad$			&		&	BB		&						&$	43.9 	_{	-1.7 	}^{+	1.7 	}$&						&	$	0.49 	^{+	0.15 	}_{-	0.13 	}$&$	0.49 	^{+	0.15 	}_{-	0.13 	}$&		3018.4 	&	$\surd$	\\
$\quad$			&		&	$\rm PL+BB$		&$	-1.59 	_{	-0.16 	}^{+	0.15 	}$&$	44.5 	_{	-1.9 	}^{+	1.9 	}$&						&	$	0.53 	^{+	0.20 	}_{-	0.16 	}$&$	0.46 	^{+	0.16 	}_{-	0.13 	}$&		3069.3 	&		\\
\hline																																												
4.6 	$\sim$	8.6 	&	31 	&	CPL		&$	0.60 	_{	-0.15 	}^{+	0.16 	}$&$	61.0 	_{	-5.7 	}^{+	5.6 	}$&						&	$	0.72 	^{+	0.41 	}_{-	0.27 	}$&						&		2936.6 	&		\\
$\quad$			&		&	Band		&$	0.66 	_{	-0.17 	}^{+	0.17 	}$&$	152.2 	_{	-7.7 	}^{+	7.6 	}$&$	-3.34 	_{	-0.41 	}^{+	0.42 	}$&	$	0.77 	^{+	0.25 	}_{-	0.19 	}$&						&		2954.9 	&		\\
$\quad$			&		&	BB		&						&$	36.6 	_{	-1.0 	}^{+	1.0 	}$&						&	$	0.63 	^{+	0.14 	}_{-	0.12 	}$&$	0.63 	^{+	0.14 	}_{-	0.12 	}$&		2900.5 	&		\\
$\quad$			&		&	$\rm PL+BB$		&$	-1.43 	_{	-0.12 	}^{+	0.15 	}$&$	36.6 	_{	-1.2 	}^{+	1.2 	}$&						&	$	0.69 	^{+	0.23 	}_{-	0.19 	}$&$	0.58 	^{+	0.16 	}_{-	0.13 	}$&		2884.2 	&	$\surd$	\\
\hline																																												
8.6 	$\sim$	30.0 	&	18 	&	CPL		&$	0.09 	_{	-0.24 	}^{+	0.24 	}$&$	36.7 	_{	-5.8 	}^{+	5.9 	}$&						&	$	0.12 	^{+	0.09 	}_{-	0.06 	}$&						&		4661.5 	&		\\
$\quad$			&		&	Band		&$	0.27 	_{	-0.32 	}^{+	0.34 	}$&$	69.5 	_{	-6.9 	}^{+	6.7 	}$&$	-3.07 	_{	-0.39 	}^{+	0.40 	}$&	$	0.15 	^{+	0.05 	}_{-	0.07 	}$&						&		4675.2 	&		\\
$\quad$			&		&	BB		&						&$	16.8 	_{	-0.8 	}^{+	0.8 	}$&						&	$	0.09 	^{+	0.04 	}_{-	0.03 	}$&$	0.09 	^{+	0.04 	}_{-	0.03 	}$&		4641.1 	&		\\
$\quad$			&		&	$\rm PL+BB$		&$	-1.63 	_{	-0.20 	}^{+	0.21 	}$&$	16.8 	_{	-1.1 	}^{+	1.1 	}$&						&	$	0.12 	^{+	0.08 	}_{-	0.05 	}$&$	0.08 	^{+	0.05 	}_{-	0.03 	}$&		4623.4 	&	$\surd$	\\
\hline																																												
30.0 	$\sim$	38.0 	&	15 	&	CPL		&$	-0.66 	_{	-0.20 	}^{+	0.20 	}$&$	101.6 	_{	-28.2 	}^{+	27.1 	}$&						&	$	0.19 	^{+	0.17 	}_{-	0.10 	}$&						&		3608.3 	&		\\
$\quad$			&		&	Band		&$	-0.56 	_{	-0.28 	}^{+	0.27 	}$&$	117.7 	_{	-26.0 	}^{+	24.7 	}$&$	-2.78 	_{	-0.48 	}^{+	0.48 	}$&	$	0.22 	^{+	0.15 	}_{-	0.11 	}$&						&		3617.6 	&		\\
$\quad$			&		&	BB		&						&$	21.3 	_{	-1.8 	}^{+	1.8 	}$&						&	$	0.12 	^{+	0.09 	}_{-	0.06 	}$&$	0.12 	^{+	0.09 	}_{-	0.06 	}$&		3610.1 	&		\\
$\quad$			&		&	$\rm PL+BB$		&$	-1.77 	_{	-0.13 	}^{+	0.13 	}$&$	25.7 	_{	-3.8 	}^{+	3.7 	}$&						&	$	0.18 	^{+	0.17 	}_{-	0.09 	}$&$	0.09 	^{+	0.14 	}_{-	0.06 	}$&		3576.8 	&	$\surd$	\\
\hline																																												
38.0 	$\sim$	42.0 	&	35 	&	CPL		&$	0.18 	_{	-0.14 	}^{+	0.14 	}$&$	49.7 	_{	-4.9 	}^{+	4.8 	}$&						&	$	0.59 	^{+	0.28 	}_{-	0.21 	}$&						&		2974.6 	&		\\
$\quad$			&		&	Band		&$	0.40 	_{	-0.17 	}^{+	0.17 	}$&$	98.5 	_{	-5.4 	}^{+	5.3 	}$&$	-2.95 	_{	-0.27 	}^{+	0.27 	}$&	$	0.68 	^{+	0.20 	}_{-	0.18 	}$&						&		2980.6 	&		\\
$\quad$			&		&	BB		&						&$	23.6 	_{	-0.6 	}^{+	0.6 	}$&						&	$	0.50 	^{+	0.10 	}_{-	0.09 	}$&$	0.50 	^{+	0.10 	}_{-	0.09 	}$&		2982.1 	&		\\
$\quad$			&		&	$\rm PL+BB$		&$	-1.59 	_{	-0.08 	}^{+	0.08 	}$&$	24.1 	_{	-0.8 	}^{+	0.8 	}$&						&	$	0.62 	^{+	0.18 	}_{-	0.15 	}$&$	0.41 	^{+	0.12 	}_{-	0.10 	}$&		2946.0 	&	$\surd$	\\
\hline																																												
42.0 	$\sim$	85.0 	&	32 	&	CPL		&$	-0.36 	_{	-0.15 	}^{+	0.15 	}$&$	43.4 	_{	-5.6 	}^{+	5.6 	}$&						&	$	0.14 	^{+	0.07 	}_{-	0.05 	}$&						&		5461.2 	&		\\
$\quad$			&		&	Band		&$	0.00 	_{	-0.21 	}^{+	0.22 	}$&$	60.6 	_{	-4.3 	}^{+	4.3 	}$&$	-2.83 	_{	-0.23 	}^{+	0.23 	}$&	$	0.17 	^{+	0.05 	}_{-	0.05 	}$&						&		5469.0 	&		\\
$\quad$			&		&	BB		&						&$	14.1 	_{	-0.4 	}^{+	0.4 	}$&						&	$	0.11 	^{+	0.03 	}_{-	0.02 	}$&$	0.11 	^{+	0.03 	}_{-	0.02 	}$&		5510.4 	&		\\
$\quad$			&		&	$\rm PL+BB$		&$	-1.73 	_{	-0.08 	}^{+	0.08 	}$&$	14.6 	_{	-0.7 	}^{+	0.7 	}$&						&	$	0.17 	^{+	0.05 	}_{-	0.04 	}$&$	0.08 	^{+	0.04 	}_{-	0.03 	}$&		5441.4 	&	$\surd$	\\
\hline																																												
85.0 	$\sim$	115.0 	&	44 	&	CPL		&$	-0.52 	_{	-0.11 	}^{+	0.11 	}$&$	39.5 	_{	-4.0 	}^{+	3.9 	}$&						&	$	0.21 	^{+	0.07 	}_{-	0.06 	}$&						&		5014.6 	&		\\
$\quad$			&		&	Band		&$	-0.16 	_{	-0.23 	}^{+	0.23 	}$&$	50.5 	_{	-3.7 	}^{+	3.7 	}$&$	-2.84 	_{	-0.21 	}^{+	0.21 	}$&	$	0.27 	^{+	0.06 	}_{-	0.09 	}$&						&		5014.7 	&		\\
$\quad$			&		&	BB		&						&$	11.9 	_{	-0.3 	}^{+	0.3 	}$&						&	$	0.17 	^{+	0.03 	}_{-	0.03 	}$&$	0.17 	^{+	0.03 	}_{-	0.03 	}$&		5147.8 	&		\\
$\quad$			&		&	$\rm PL+BB$		&$	-1.81 	_{	-0.06 	}^{+	0.06 	}$&$	12.4 	_{	-0.5 	}^{+	0.5 	}$&						&	$	0.27 	^{+	0.06 	}_{-	0.05 	}$&$	0.10 	^{+	0.04 	}_{-	0.03 	}$&		4998.4 	&	$\surd$	\\
\hline																																												
115.0 	$\sim$	125.0 	&	68 	&	CPL		&$	-0.88 	_{	-0.07 	}^{+	0.07 	}$&$	61.1 	_{	-5.2 	}^{+	5.3 	}$&						&	$	0.56 	^{+	0.12 	}_{-	0.11 	}$&						&		3847.6 	&		\\
$\quad$			&		&	Band		&$	-0.69 	_{	-0.14 	}^{+	0.15 	}$&$	60.3 	_{	-4.4 	}^{+	4.4 	}$&$	-2.85 	_{	-0.25 	}^{+	0.25 	}$&	$	0.64 	^{+	0.15 	}_{-	0.15 	}$&						&		3845.2 	&	$\surd$	\\
$\quad$			&		&	BB		&						&$	12.6 	_{	-0.2 	}^{+	0.2 	}$&						&	$	0.43 	^{+	0.06 	}_{-	0.05 	}$&$	0.43 	^{+	0.06 	}_{-	0.05 	}$&		4351.1 	&		\\
$\quad$			&		&	$\rm PL+BB$		&$	-1.87 	_{	-0.04 	}^{+	0.04 	}$&$	14.1 	_{	-0.6 	}^{+	0.6 	}$&						&	$	0.70 	^{+	0.11 	}_{-	0.09 	}$&$	0.22 	^{+	0.08 	}_{-	0.06 	}$&		3859.0 	&		\\
\hline																																												
125.0 	$\sim$	138.0 	&	26 	&	CPL		&$	-0.84 	_{	-0.22 	}^{+	0.22 	}$&$	35.6 	_{	-7.5 	}^{+	7.3 	}$&						&	$	0.16 	^{+	0.09 	}_{-	0.08 	}$&						&		4057.4 	&	$\surd$	\\
$\quad$			&		&	Band		&$	-0.65 	_{	-0.28 	}^{+	0.28 	}$&$	37.0 	_{	-2.9 	}^{+	2.9 	}$&$	-3.10 	_{	-0.33 	}^{+	0.33 	}$&	$	0.19 	^{+	0.01 	}_{-	0.07 	}$&						&		4069.0 	&		\\
$\quad$			&		&	BB		&						&$	9.0 	_{	-0.3 	}^{+	0.3 	}$&						&	$	0.12 	^{+	0.04 	}_{-	0.03 	}$&$	0.12 	^{+	0.04 	}_{-	0.03 	}$&		4098.8 	&		\\
$\quad$			&		&	$\rm PL+BB$		&$	-2.11 	_{	-0.11 	}^{+	0.11 	}$&$	10.0 	_{	-0.8 	}^{+	0.8 	}$&						&	$	0.17 	^{+	0.08 	}_{-	0.05 	}$&$	0.06 	^{+	0.06 	}_{-	0.03 	}$&		4074.0 	&		\\

\hline
\end{tabular}
\begin{tablenotes}
\item[a]The spectral index of Band, CPL, and PL models.
\item[b]The peak energy, cut energy, Blackbody temperature of Band, CPL, and BB models, respectively, in units of keV.
\item[c]The observed total flux and BB flux, respectively. The flux is in units of $\rm 10^{-6}~erg~cm^{-2}~s^{-1}$.
\end{tablenotes}
\end{threeparttable} 
\label{table1}
\end{table*}

\clearpage
\begin{table*} \centering
\renewcommand\tabcolsep{2.2pt}
\renewcommand\arraystretch{1.4}
\caption{The derived parameters of pure fireball model and hybrid model.}
\begin{threeparttable} 
\begin{tabular}{cccccccccc} 
\hline\hline
$t_{1}-t_{2}$(s)&$F_{\rm BB}/F_{\rm tot}$&$\Re^{a}$&$\Gamma_{\rm ph}^{b}$&$R_{\rm ph}^{c}$&$R_{0}^{d}$&$1+\sigma_{0}^{e}$&$\eta^{e}$&$\Gamma_{\rm ph}^{f}$&$R_{\rm ph}^{g}$\\
\hline
0.0 	$\sim$	4.6 	&$	1.00 					$&$	3.6 	_{-	0.4 	}^{+	0.5 	}$&$	333.6 	_{-	18.1 	}^{+	21.4 	}$&$	19.2 	_{-	1.9 	}^{+	2.2 	}$&$	8.0 	_{-	3.7 	}^{+	4.4 	}$&	1.0 	&	797.8 	&	261.4 	&	4.9 	\\
4.6 	$\sim$	8.6 	&$	0.83 	_{-	0.29 	}^{+	0.17 	}$&$	5.6 	_{-	0.7 	}^{+	0.8 	}$&$	326.0 	_{-	24.3 	}^{+	29.7 	}$&$	29.1 	_{-	3.5 	}^{+	4.1 	}$&$	9.4 	_{-	5.8 	}^{+	7.0 	}$&	1.0 	&	779.6 	&	255.4 	&	7.5 	\\
8.6 	$\sim$	30.0 	&$	0.63 	_{-	0.39 	}^{+	0.37 	}$&$	9.7 	_{-	2.4 	}^{+	3.3 	}$&$	184.1 	_{-	23.5 	}^{+	33.1 	}$&$	28.4 	_{-	6.2 	}^{+	8.5 	}$&$	10.6 	_{-	10.6 	}^{+	16.9 	}$&	1.0 	&	440.3 	&	144.2 	&	7.3 	\\
30.0 	$\sim$	38.0 	&$	0.47 	_{-	0.43 	}^{+	0.53 	}$&$	4.4 	_{-	2.1 	}^{+	3.7 	}$&$	247.9 	_{-	43.3 	}^{+	77.9 	}$&$	17.5 	_{-	6.6 	}^{+	11.8 	}$&$	3.2 	_{-	3.2 	}^{+	11.2 	}$&	1.0 	&	593.0 	&	194.3 	&	4.5 	\\
38.0 	$\sim$	42.0 	&$	0.66 	_{-	0.23 	}^{+	0.28 	}$&$	10.9 	_{-	1.6 	}^{+	1.8 	}$&$	268.3 	_{-	19.0 	}^{+	22.4 	}$&$	46.7 	_{-	5.8 	}^{+	6.7 	}$&$	13.1 	_{-	8.2 	}^{+	9.9 	}$&	1.0 	&	641.7 	&	210.2 	&	12.0 	\\
42.0 	$\sim$	85.0 	&$	0.45 	_{-	0.20 	}^{+	0.25 	}$&$	12.7 	_{-	2.6 	}^{+	3.3 	}$&$	186.3 	_{-	15.1 	}^{+	19.0 	}$&$	37.7 	_{-	6.3 	}^{+	7.8 	}$&$	8.5 	_{-	6.9 	}^{+	9.0 	}$&	1.0 	&	445.6 	&	146.0 	&	9.7 	\\
85.0 	$\sim$	115.0 	&$	0.38 	_{-	0.14 	}^{+	0.18 	}$&$	20.4 	_{-	3.6 	}^{+	4.4 	}$&$	185.3 	_{-	11.9 	}^{+	14.4 	}$&$	60.5 	_{-	8.5 	}^{+	10.3 	}$&$	10.9 	_{-	7.5 	}^{+	9.5 	}$&	1.0 	&	443.1 	&	145.1 	&	15.5 	\\

\hline
\end{tabular}
\begin{tablenotes}
\item[a]The function $\Re \equiv (\frac{F_{\rm BB}}{\sigma T^{4}})^{1/2}$, in units of $\rm 10^{-19}$.
\item[b]The the Lorentz factor at photosphere radius derived by pure fireball model.
\item[c]The photosphere radius derived by pure fireball model, in units of $\rm 10^{11}$ cm.
\item[d]The central engine radius derived by pure fireball model, in units of $\rm 10^{7}$ cm.
\item[e]The magnetization factor and dimensionless entropy of central engine derived by hybrid model.
\item[f]The the Lorentz factor at photosphere radius derived by hybrid model.
\item[g]The photosphere radius derived by hybrid model, in units of $\rm 10^{11}$ cm.
\end{tablenotes}
\end{threeparttable} 
\label{table2}
\end{table*}

\appendix
\section{Method to derive the parameters of hybrid model}
In this appendix, we present the details and expressions of the regimes II, III, VI and V of “top-down” approach.

For regime II (see also Equation (37) in \citealt{2015ApJ...801...2}, one has:
\begin{align}
\begin{split}
& 1+\sigma_{0}=25.5(1+z)^{4/3} \left(\frac{kT_{\rm obs}}{50 \rm keV }\right)^{4/3} \times \left(\frac{F_{\rm BB}}{10^{-8} \rm erg s^{-1} cm^{-2}}\right)^{-1/3}r^{2/3}_{0,9}f^{-1}_{\rm th,-1}f^{-1}_{\gamma}d^{-2/3}_{L,28},\\
& \eta =74.8(1+z)^{11/12} \left(\frac{kT_{\rm obs}}{50 \rm keV }\right)^{11/12} \times \left(\frac{F_{\rm BB}}{10^{-8} \rm erg s^{-1} cm^{-2}}\right)^{1/48}r^{5/24}_{0,9}d^{1/24}_{L,28},\\
& R_{\rm ph} =1.78\times10^{10}{\rm cm}(1+z)^{-25/12} \left(\frac{kT_{\rm obs}}{50 \rm keV }\right)^{-25/12} \times \left(\frac{F_{\rm BB}}{10^{-8} \rm erg s^{-1} cm^{-2}}\right)^{37/48}r^{-7/24}_{0,9}d^{37/24}_{L,28},\\
& \Gamma_{\rm ph} =46.4(1+z)^{-1/12} \left(\frac{kT_{\rm obs}}{50 \rm keV }\right)^{-1/12} \times \left(\frac{F_{\rm BB}}{10^{-8} \rm erg s^{-1} cm^{-2}}\right)^{13/48}r^{-7/24}_{0,9}d^{13/24}_{L,28},\\
\label{eq:regime II}
\end{split}
\end{align}

For regime III and regime VI (see also Eq.(38) in \citealt{2015ApJ...801...2}):
\begin{align}
\begin{split}
& 1+\sigma_{0}=5.99(1+z)^{4/3} \left(\frac{kT_{\rm obs}}{50 \rm keV }\right)^{4/3} \times \left(\frac{F_{\rm BB}}{10^{-8} \rm erg s^{-1} cm^{-2}}\right)^{-1/3}r^{2/3}_{0,9}f^{-1}_{\rm th,-1}f^{-1}_{\gamma}d^{-2/3}_{L,28},\\
& \eta =20.3(1+z)^{-5/6} \left(\frac{kT_{\rm obs}}{50 \rm keV }\right)^{11/12} \times \left(\frac{F_{\rm BB}}{10^{-8} \rm erg s^{-1} cm^{-2}}\right)^{11/24}r^{-2/3}_{0,9}f^{-3/4}_{\rm th,-1}f^{-3/4}_{\gamma}d^{11/12}_{L,28},\\
& R_{\rm ph} =4.09\times10^{11}{\rm cm}(1+z)^{-3/2} \left(\frac{kT_{\rm obs}}{50 \rm keV }\right)^{-5/8} \times \left(\frac{F_{\rm BB}}{10^{-8} \rm erg s^{-1} cm^{-2}}\right)^{5/8}f^{-1/4}_{\rm th,-1}f^{-1/4}_{\gamma}d^{5/4}_{L,28},\\
\label{eq:regime III}
\end{split}
\end{align}

For regime V (see also Eq.(39) in \citealt{2015ApJ...801...2}):
\begin{align}
\begin{split}
& 1+\sigma_{0}=6.43(1+z)^{4/3} \left(\frac{kT_{\rm obs}}{50 \rm keV }\right)^{4/3} \times \left(\frac{F_{\rm BB}}{10^{-8} \rm erg s^{-1} cm^{-2}}\right)^{-1/3}r^{2/3}_{0,9}f^{-1}_{\rm th,-1}f^{-1}_{\gamma}d^{-2/3}_{L,28},\\
& \eta =105.0(1+z)^{7/6} \left(\frac{kT_{\rm obs}}{50 \rm keV }\right)^{7/6} \times \left(\frac{F_{\rm BB}}{10^{-8} \rm erg s^{-1} cm^{-2}}\right)^{5/24}r^{1/12}_{0,9}f^{1/2}_{\rm th,-1}f^{1/2}_{\gamma}d^{5/12}_{L,28},\\
& R_{\rm ph} =4.62\times10^{10}{\rm cm}(1+z)^{-13/6} \left(\frac{kT_{\rm obs}}{50 \rm keV }\right)^{-13/6} \times \left(\frac{F_{\rm BB}}{10^{-8} \rm erg s^{-1} cm^{-2}}\right)^{17/24}r^{-1/4}_{0,9}f^{-1/6}_{\rm th,-1}f^{-1/6}_{\gamma}d^{17/12}_{L,28},\\
& \Gamma_{\rm ph} =15.3(1+z)^{-1/6} \left(\frac{kT_{\rm obs}}{50 \rm keV }\right)^{-1/6} \times \left(\frac{F_{\rm BB}}{10^{-8} \rm erg s^{-1} cm^{-2}}\right)^{-5/24}r^{-1/4}_{0,9}f^{-1/6}_{\rm th,-1}f^{-1/6}_{\gamma}d^{5/24}_{L,28},\\
\label{eq:regime V}
\end{split}
\end{align}

One needs to note that regime VI has the identical scalings as regime III. The ratio $f_{\gamma}$=$L_{\gamma}/L_{\rm w}$ ($f_{\gamma}$=0.5 we adopted in this paper) relates the gamma-ray luminosity (derived from $F_{\rm tot}$) to the wind luminosity $L_{\rm w}$, and $f_{\rm th}$=$F_{\rm BB}$/$F_{\rm obs}$ which is the thermal flux ratio can be directly measured from the data. The constants $r_{0}$ are set equal to the $R_{0}$ values in Section 3.
\end{document}